\documentclass{jfm}
\usepackage{graphicx}
\usepackage{newtxtext}
\usepackage{newtxmath}
\usepackage{natbib}
\usepackage[english]{babel}
\usepackage[utf8]{inputenc}
\usepackage{amsmath}
\usepackage{hyperref}
\hypersetup{
    colorlinks = true,
    urlcolor   = blue,
    citecolor  = black,
    linkcolor=blue,
}
\usepackage[table,xcdraw]{xcolor}
\usepackage{tabularx}
\usepackage{amsmath}
\usepackage{bigints}
\newcommand{\RomanNumeralCaps}[1]
\linenumbers

\title{Analytical solution for the cumulative wake of wind turbines in wind farms}

\author{Majid Bastankhah\aff{1}
  \corresp{\email{majid.bastankhah@durham.ac.uk}},
  Bridget L. Welch\aff{1},
  Luis A. Mart\'{i}nez-Tossas\aff{2},
  Jennifer King \aff{2} \\
 \and Paul Fleming\aff{2}}

\affiliation{\aff{1}Department of Engineering, Durham University, Durham DH1 3LE, United Kingdom
\aff{2}National Wind Technology Center, National Renewable Energy Laboratory, Golden, CO 80401 USA}
\usepackage{xcolor}
\begin{document}

\maketitle

\begin{abstract}
\color{black}
This paper solves an approximate form of conservation of mass and momentum for a turbine in a wind farm array. The solution is a fairly simple explicit relationship that predicts the streamwise velocity distribution within a wind farm with an arbitrary layout. \color{black} As this model is obtained by solving flow governing equations directly for a turbine that is subject to upwind turbine wakes, no ad hoc superposition technique is needed to predict wind farm flows. \color{black} A suite of large-eddy simulations (LES) of wind farm arrays is used to examine self-similarity as well as validity of the so-called conservation of momentum deficit for turbine wakes in wind farms. The simulations are performed with and without the presence of some specific turbines in the wind farm. This allows us to systematically study some of the assumptions made to develop the analytical model. A modified version of the conservation of momentum deficit is also proposed to provide slightly better results at short downwind distances, as well as in the far wake of turbines deep inside a wind farm. Model predictions are validated against the LES data for turbines in both full-wake and partial-wake conditions. While our results highlight the limitation in capturing the flow speed-up between adjacent turbine columns, the model is overall able to acceptably predict flow distributions for \color{black}a moderately \color{black} sized wind farm.
\color{black} Finally, the paper employs the new model to provide insights on the accuracy of common wake superposition methods.
\end{abstract}

\section{Introduction}\label{sec:intro}
Fast running engineering (i.e., control-oriented) wake models  are arguably still the most popular tools in the wind energy industry for the design and active control of wind farms due to their simplicity and low computational cost \citep{stevens2017review,porte2020}. 
In order to improve wind farm flow modelling without increasing computational costs, we need to develop a better theoretical understanding of wind flow physics in wind farms \citep{meneveau2019big}. Modelling of wind farm flows based on engineering wake models often consists of two steps: (i) modelling wakes of each individual wind turbine separately and then (ii) using wake superposition techniques to take into account cumulative wake effects in wind farms. The most common superposition techniques are \emph{linear} \citep{Lissaman1979} and \emph{root-sum-square} \citep{Katic1987}, denoted by A.I and A.II in table \ref{Table}, respectively. For each method, the table shows how the total velocity deficit $\Delta U$ at a given position depends on the velocity deficit $\Delta U_i$ caused by the i$^{th}$ wind turbine (WT$_i$), where $i$ changes from 1 to $n$, and $n$ is the number of wind turbines upstream of that position in a wind farm. In addition, different methods are used in the literature to compute $\Delta U_i$, as shown in table \ref{Table}. By definition, the velocity deficit for WT$_i$ is equal to the difference between the incoming velocity and the wake velocity denoted by $U_i$. The incoming velocity, however, can be defined either based on (i) the velocity at the inlet of the wind farm denoted by $U_{0}$ (method B.I) \citep{Lissaman1979,Katic1987}, or (ii) the local incoming velocity for WT$_{i}$ denoted by $U_{in, i}$ \citep{voutsinas1990,Niayifar2016}.      

\begin{center}
 \begin{table}

\begin{tabularx}{1\linewidth}{ m{.25\linewidth} | m{.25\linewidth} | m{.25\linewidth} | m{.25\linewidth}}

         \multicolumn{2}{c|}{\textbf{A.} Superposition method}                                                
         & \multicolumn{2}{c}{\textbf{B.} Incoming velocity}                \\ \hline
         
  \textbf{A.I.} Linear & \textbf{A.II.} Root-sum-square & \textbf{B.I.} Global &  \textbf{B.II.} Local \\ \hline
 $\Delta U=\sum\limits_{i=1}^{n}\Delta U_{i}$& 	$\Delta U=\sqrt{\sum\limits_{i=1}^{n}\Delta U_{i}^2}$ & $\Delta U_i=U_{0}-U_i$ & $\Delta U_i=U_{in, i}-U_i$\\
   \end{tabularx}
  \caption{Wake superposition methods commonly used in the literature to estimate cumulative wake effects in wind farms. }
  \label{Table}
\end{table}
\end{center}

These different superposition methods are often claimed to conserve flow properties. For instance, the \emph{linear} superposition method (A.I) is perceived to conserve \emph{momentum deficit} \citep{Lissaman1979}, while the \emph{root-sum-square} method (A.II) is thought to conserve the \emph{energy deficit} \citep{Katic1987}. However, as described later in \S\ref{sec:stw}, the validity of these relationships has not been proven, and so these superposition methods should be viewed as empirical relationships. A result-driven approach was mostly adopted to develop such methods. 
For instance, a superposition method consisting of methods A.I and B.I used by \citet{Lissaman1979} is known to result in negative velocity values in large wind farms. \citet{Katic1987} overcame this issue by proposing a new method including A.II in conjunction with B.I.  More recently, \citet{zong2020} have developed a more physics-based superposition method based on the mean convection velocity for each turbine wake. The convection velocity for the combined wake is obtained through an iterative approach. Other recent studies have examined the accuracy of existing superposition models \citep[e.g.,][ among others]{vogel2020,stallard2016}. Overall, there is not unanimous agreement in the literature on which method is most accurate over the wide variety of operating conditions of wind farms.

In this study, an approach that does not rely on the superposition of single turbine wake models is adapted based on a holistic view of turbines in a wind farm. Unlike prior studies that assume each turbine can be treated as a single turbine and separated from the rest of the wind farm, we directly solve governing equations for wind turbines within a wind farm. This eliminates the need to introduce any superposition method. In the following, \color{black}\S\ref{sec:integral_form_RANs} develops the integral form of governing equations for turbine wake flows in wind farms. The large eddy simulation (LES) setup is described in \S\ref{sec:LES_setup}. \S\ref{sec:model_derivation} derives the analytical solution of conservation of momentum deficit for a turbine within a wind farm. Model predictions are presented in \S\ref{sec:model_predictions}. A modified version of the conservation of momentum deficit is proposed and solved in \S\ref{sec:modified_momentum_deficit}. In \S\ref{sec:stw}, we examine common wake superposition techniques in the literature. Finally, a summary and a discussion about possible future research directions are provided in \S\ref{sec:summary}.

\section{Integral form of governing equations for turbine wakes within a wind farm}\label{sec:integral_form_RANs}
\begin{figure}
	\begin{center}
		\includegraphics[width=1\textwidth]{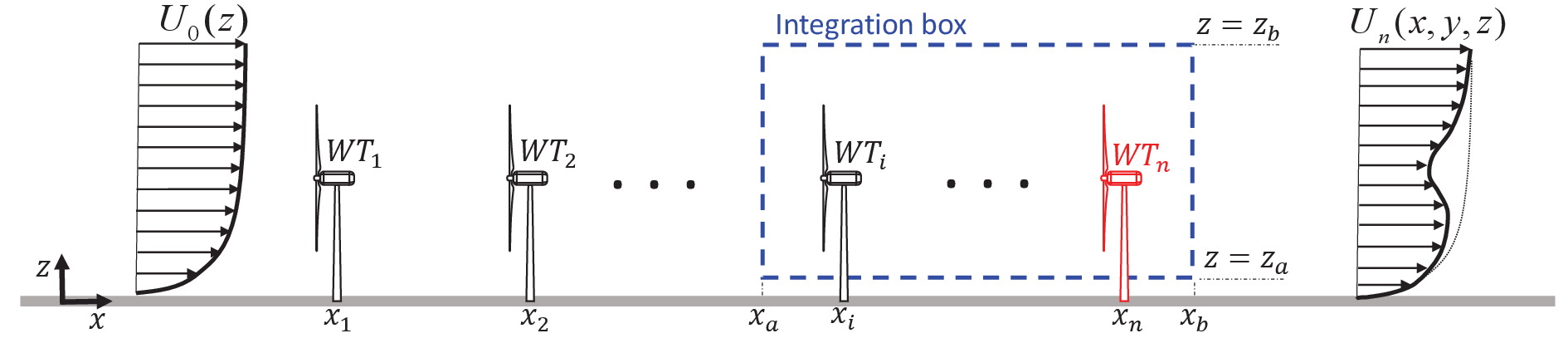}
		\caption{\color{black} Schematic of a wind farm with an arbitrary layout consisting of $n$ wind turbines (WT$_1$,WT$_2$, ... WT$_n$) immersed in a turbulent boundary layer flow. The momentum equation (\ref{Eq:Nav-Stk_general-x3}) is integrated over the shown box. The integration is performed with and without WT$_n$, shown in the figure by the red colour. }	
		\label{Fig:schematic_CV}
	\end{center}
\end{figure}

\color{black}
Let us assume a wind farm with an arbitrary layout of $n$ wind turbines (WT$_1$,WT$_2$, ... WT$_i$, ... WT$_n$) immersed in a turbulent boundary layer flow with a velocity profile denoted by $U_0$. The position of WT$_{i}$ is denoted by $\mathbf{X}_i=(x_i,y_i,z_i)$, where $x$, $y$ and $z$ are the streamwise, spanwise, and vertical directions in the coordinate system, respectively. Turbines are labelled with respect to their streamwise positions such that $x_i\geq x_{i-1}$, where $i=\{2,3,$ ... $ n\}$. The Reynolds-averaged Navier-Stokes (RANS) equation in the streamwise direction at high Reynolds numbers (neglecting viscosity effects) is given by \citep{shapiro2018,shapiro2019paradigm}
\begin{equation}\label{Eq:Nav-Stk_general-x}
U\frac{\partial U}{\partial x}+  V\frac{\partial U}{\partial y} +   W\frac{\partial U}{\partial z} =-\frac{1}{\rho}\frac{\partial P}{\partial x}- \frac{\partial \overline{u^2}}{\partial x}- \frac{\partial \overline{uv}}{\partial y} -  \frac{\partial \overline{uw}}{\partial z}+\sum_{i=1}^n f_i,
\end{equation}
where $U$, $V$ and $W$ are the time-averaged streamwise ($x$), lateral ($y$) and vertical ($z$) velocity components, respectively. Turbulent velocity fluctuations are represented by $u$, $v$ and $w$ and the overbar denotes time averaging. Also, $P$ is the time-averaged static pressure and $\rho$ is the air density. The term $f_i$ represents the effect of the thrust force of WT$_i$ on the horizontal momentum and is given by
\begin{equation}
    f_i=-{T_i}/(\rho \pi R^2)\delta\left(x-x_i\right)H\left(R^2-\left[\left(y-y_i\right)^2+\left(z-z_i\right)^2\right]\right),
\end{equation}
where $T_i$ is the magnitude of the thrust force of WT$_i$ in the streamwise direction, $R$ is the turbine radius, $\delta(x)$ is the Dirac delta function, and $H(x)$ is the Heaviside step function. Using the incoming boundary-layer profile $U_0(z)$, (\ref{Eq:Nav-Stk_general-x}) can be written as
\begin{multline}\label{Eq:Nav-Stk_general-x2}
U\frac{\partial (U_0-U)}{\partial x}+  V\frac{\partial(U_0- U)}{\partial y} +   W\frac{\partial(U_0- U)}{\partial z} =  \frac{1}{\rho}\frac{\partial P}{\partial x}+ \frac{\partial \overline{u^2}}{\partial x}+ \frac{\partial \overline{uv}}{\partial y} + \frac{\partial \overline{uw}}{\partial z}+ W\frac{\textrm{d} U_{0}}{\textrm{d} z}-\sum_{i=1}^n f_i.
\end{multline}
From the continuity equation, we know
\begin{equation}\label{Eq:continuity}
    \frac{\partial U}{\partial x}+\frac{\partial V}{\partial y}+\frac{\partial W}{\partial z}=0.
\end{equation}
Multiplying (\ref{Eq:continuity}) by $(U_0-U)$ and adding the product to the left-hand side of  (\ref{Eq:Nav-Stk_general-x2}) yields
\begin{multline}\label{Eq:Nav-Stk_general-x3}
\frac{\partial U(U_0-U)}{\partial x}+  \frac{\partial V(U_0- U)}{\partial y} +   \frac{\partial W(U_0- U)}{\partial z} = \frac{1}{\rho}\frac{\partial P}{\partial x}+ \frac{\partial \overline{u^2}}{\partial x}+ \frac{\partial \overline{uv}}{\partial y} +  \frac{\partial \overline{uw}}{\partial z}+ W\frac{\textrm{d} U_{0}}{\textrm{d} z}-\sum_{i=1}^n f_i.
\end{multline}
Next, (\ref{Eq:Nav-Stk_general-x3}) is integrated from $x_a$ to $x_b$ with respect to $x$, from $y_a$ to $y_b$ with respect to $y$ and from $z_a$ to $z_b$ with respect to $z$, where $x_a\ll x_n < x_b$, $y_a\ll y_n \ll y_b$ and $0\ll z_a< z_n \ll z_b$. Note that $z_a\gg 0$ to ensure that the assumption of negligible viscous forces is valid. The value of $z_a$ can be equal to zero only if the Reynolds shear stresses in (\ref{Eq:Nav-Stk_general-x3}) are replaced with the total shear stresses (i.e., sum of turbulent and viscous shear stresses). Without loss of generality, we assume that the integration box includes WT$_n$ and an arbitrary number of upwind turbines as shown in figure \ref{Fig:schematic_CV}. Integrating (\ref{Eq:Nav-Stk_general-x3}) yields
\begin{multline}\label{Eq:Nav-Stk_general-x5a}
\sum_{i\in B}\frac{ T_i}{\rho}=\int U_n(U_0-U_n)\textrm{d}A\bigg\rvert^{x=x_b}_{x=x_a}- \frac{1}{\rho}\int P_n\textrm{d}A\bigg\rvert^{x=x_b}_{x=x_a}\\- \int \overline{u_n^2}\textrm{d}A\bigg\rvert^{x=x_b}_{x=x_a}-\int\overline{uw}_{n}\textrm{d}x\textrm{d}y\bigg\rvert^{z=z_b}_{z=z_a}-\int \frac{\textrm{d} U_0}{\textrm{d} z}W_{n}\textrm{d}V,
\end{multline}
where $i$ is a member of set $B$ if WT$_i$ is inside the integration box. Also $\textrm{d}A$ is $\textrm{d}y\textrm{d}z$ and $\textrm{d}V$ is $\textrm{d}x\textrm{d}y\textrm{d}z$. In (\ref{Eq:Nav-Stk_general-x5a}) and hereafter, any velocity or pressure term with a subscript $i$ denotes the value of the given variable in the presence of WT$_1$,WT$_2$, ... WT$_i$. Now, we perform the same integration once more but this time in the absence of WT$_n$. This leads to
\begin{multline}\label{Eq:Nav-Stk_general-x5b}
\sum_{i\in B'}\frac{ T_i}{\rho}=\int U_{n-1}(U_0-U_{n-1})\textrm{d}A\bigg\rvert^{x=x_b}_{x=x_a}- \frac{1}{\rho}\int P_{n-1}\textrm{d}A\bigg\rvert^{x=x_b}_{x=x_a}\\- \int \overline{u_{n-1}^2}\textrm{d}A\bigg\rvert^{x=x_b}_{x=x_a}-\int\overline{uw}_{n-1}\textrm{d}x\textrm{d}y\bigg\rvert^{z=z_b}_{z=z_a}-\int \frac{\textrm{d} U_0}{\textrm{d} z}W_{n-1}\textrm{d}V,
\end{multline}
where set $B'$ is equal to set $B$ excluding $n$ (i.e., $B\setminus\{n\}=\{i:i\in B$ and $i\not\in \{n\}\}$). As $x_a\ll x_n$, surface integrals at $x=x_a$ provide the same results in both (\ref{Eq:Nav-Stk_general-x5a}) and (\ref{Eq:Nav-Stk_general-x5b}). By subtracting (\ref{Eq:Nav-Stk_general-x5b}) from (\ref{Eq:Nav-Stk_general-x5a}), we obtain
\begin{multline}\label{Eq:Nav-Stk_general-x5}
\underbrace{\frac{ T_n}{\rho}}_\text{Thrust}=\underbrace{\left[\int U_n(U_0-U_n)\textrm{d}A-\int U_{n-1}(U_0-U_{n-1})\textrm{d}A\right]}_\text{Momentum deficit}\underbrace{- \frac{1}{\rho}\int \left(P_n-P_{n-1}\right)\textrm{d}A}_\text{Pressure}\\\underbrace{- \int \left(\overline{u_n^2}-\overline{u_{n-1}^2}\right)\textrm{d}A}_\text{Reynolds normal stress}\underbrace{-\int\left(\overline{uw}_{n}-\overline{uw}_{n-1}\right)\bigg\rvert^{z=z_b}_{z=z_a}\textrm{d}x\textrm{d}y}_\text{Reynolds shear stress} \underbrace{-\int \frac{\textrm{d} U_0}{\textrm{d} z}\left(W_{n}-W_{n-1}\right)\textrm{d}V}_\text{Mean flow shear},
\end{multline}
where $\textrm{d}A$ in (\ref{Eq:Nav-Stk_general-x5}) is $\textrm{d}y\textrm{d}z$ at $x=x_b$. Note that the convective terms, which include cross-stream velocity components, as well as the lateral Reynolds shear stress term in (\ref{Eq:Nav-Stk_general-x3}), vanish in equations written after (\ref{Eq:Nav-Stk_general-x3}), according to the fundamental theorem of calculus. For instance, $\int_{y_a}^{y_b}\frac{\partial V(U_0- U)}{\partial y}\textrm{d}y=V(U_0-U)\big\rvert^{y=y_b}_{y=y_a}$, and therefore the difference in this term with and without WT$_n$ is expected to be zero for $y_a\ll y_n \ll y_b$ \citep[][]{tennekes1972first,pope2000}. However, the term including $\overline{uw}$ is kept due to the presence of the boundary-layer flow. LES data are used in the next section to quantify the value of each term in (\ref{Eq:Nav-Stk_general-x5}) for wind turbines in a wind farm.

\section{Large eddy simulation (LES) setup}\label{sec:LES_setup}
\begin{figure}
	\begin{center}
		\includegraphics[width=.8\textwidth]{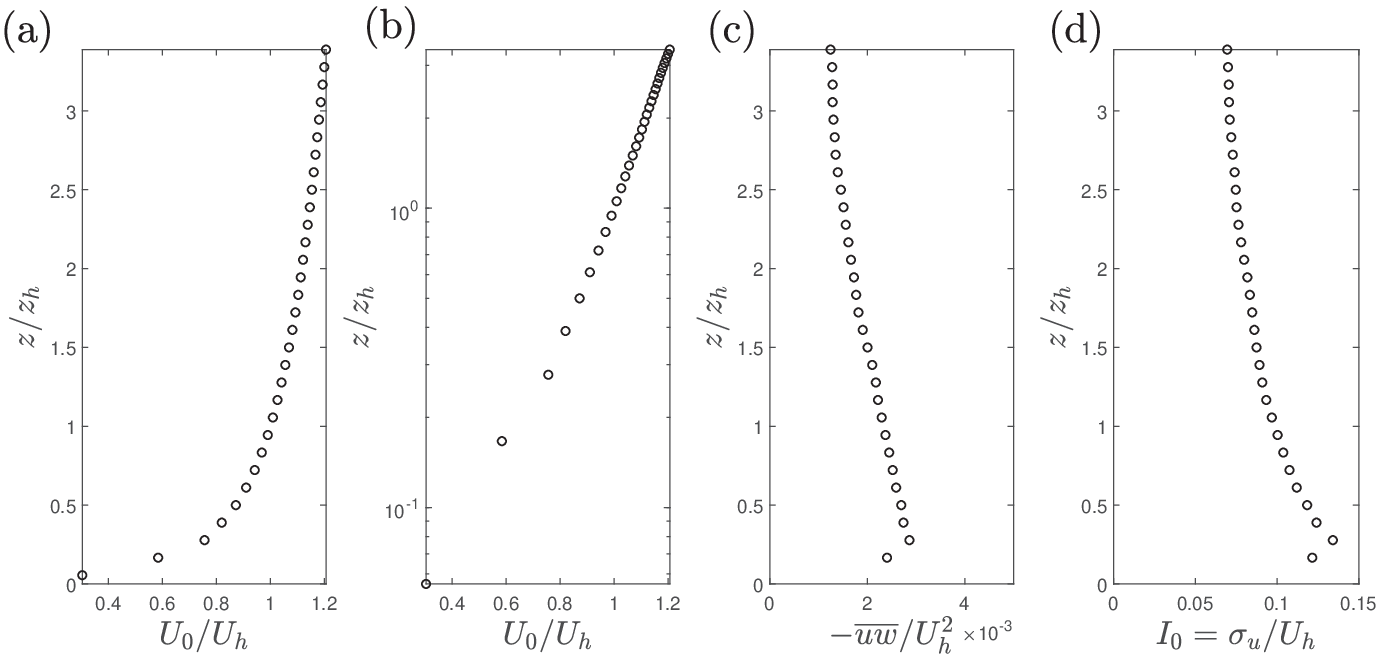}
		\caption{\color{black} Spanwise-averaged vertical profiles of inflow  characteristics obtained from precursor simulations: (a) the normalised streamwise velocity $U_0/U_h$, where $U_h=U_0(z=z_h)$, in a linear scale, (b) the normalised streamwise velocity $U_0/U_h$ in a semi-logarithmic scale, (c) the normalised Reynolds shear stress $-\overline{uw}/U_h^2$, and (d) the incoming turbulence intensity $I_0=\sigma_u/U_h$.}	
		\label{Fig:inflow}
	\end{center}
\end{figure}

\color{black} 
We use results from LES to compare against predictions of the analytical model presented in this paper. The LES is performed using the Simulator fOr Applications (SOWFA). SOWFA solves the incompressible filtered Navier-Stokes equations using a finite-volume formulation \citep{churchfield2012}. 
A precursor simulation is first performed to generate the inflow boundary conditions for the simulation with the turbines. The precursor simulation uses a 5 [km] by 5 [km] by 1 [km] domain with 10 [m] resolution in all directions.
The desired velocity of 8 [m/s] at turbine hub height ($z_h=$ 90 [m]) is achieved by  adjusting the pressure gradient at every time-step \citep{churchfield2012}.
A neutral boundary layer is simulated with a capping inversion at 750 [m].
\color{black} A wall-stress model with a roughness height of $z_0=0.15$ [m] was used to represent the shear stresses at the wall \citep{schumann1975subgrid,grotzbach1987direct}\color{black}.
The simulation is run for 20,000 [s] of simulated time for the turbulence to develop and then data on a boundary is sampled for 5,000 [s] to use in the simulation with the turbines as inflow.
Figure \ref{Fig:inflow} shows the spanwise averaged velocity and turbulence profiles as a function of height for the precursor simulation. All simulations presented use the same precursor simulation\color{black}. A slight logarithmic layer mismatch is observed in the surface layer shown in figure \ref{Fig:inflow}b, which is common in LES of atmospheric boundary layers (ABLs) \citep{brasseur2010}.\color{black}  

\begin{figure}
	\begin{center}
		\includegraphics[width=.7\textwidth]{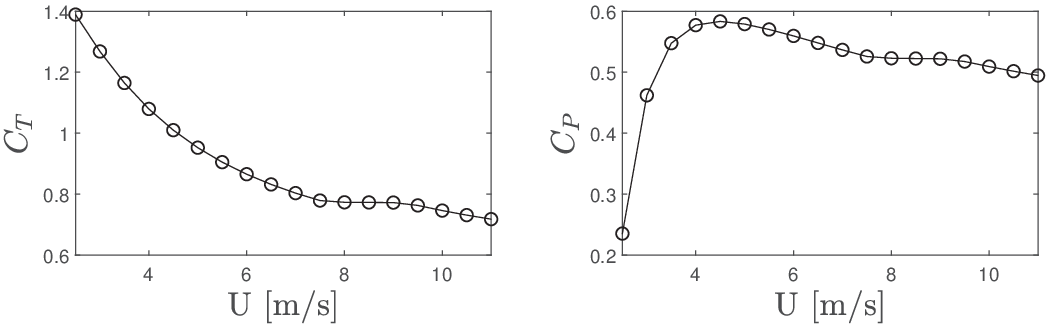}
		\caption{\color{black} Thrust and power coefficient of the NREL 5MW actuator disk model with rotation.}	
		\label{Fig:ct_cp}
	\end{center}
\end{figure}

The simulations with the turbines use the precursor as an inflow boundary condition and initial condition.
The boundary conditions on the sides are periodic, and the outflow boundary condition adjusts the velocity field to conserve mass.
The domain is a sub-sample of the precursor domain with 5 [km] by 1.8 [km] by 1 [km] and 10 [m] grid resolution in all directions.
The turbines are modelled using an actuator disk with rotation \citep{martinez2015}. The blade aerodynamic properties are from the NREL 5MW wind turbine \citep{jonkman2009definition}. 
A conventional variable-speed and variable blade-pitch-to-feather control system is used to control the turbine  \citep{jonkman2009definition}.
The power and thrust coefficients for the turbine were determined by running a set of simulations with uniform inflow at different wind speeds.
Figure \ref{Fig:ct_cp} shows the thrust and power coefficients for the turbine model used in the simulations.

\begin{figure}
	\begin{center}
		\includegraphics[width=1\textwidth]{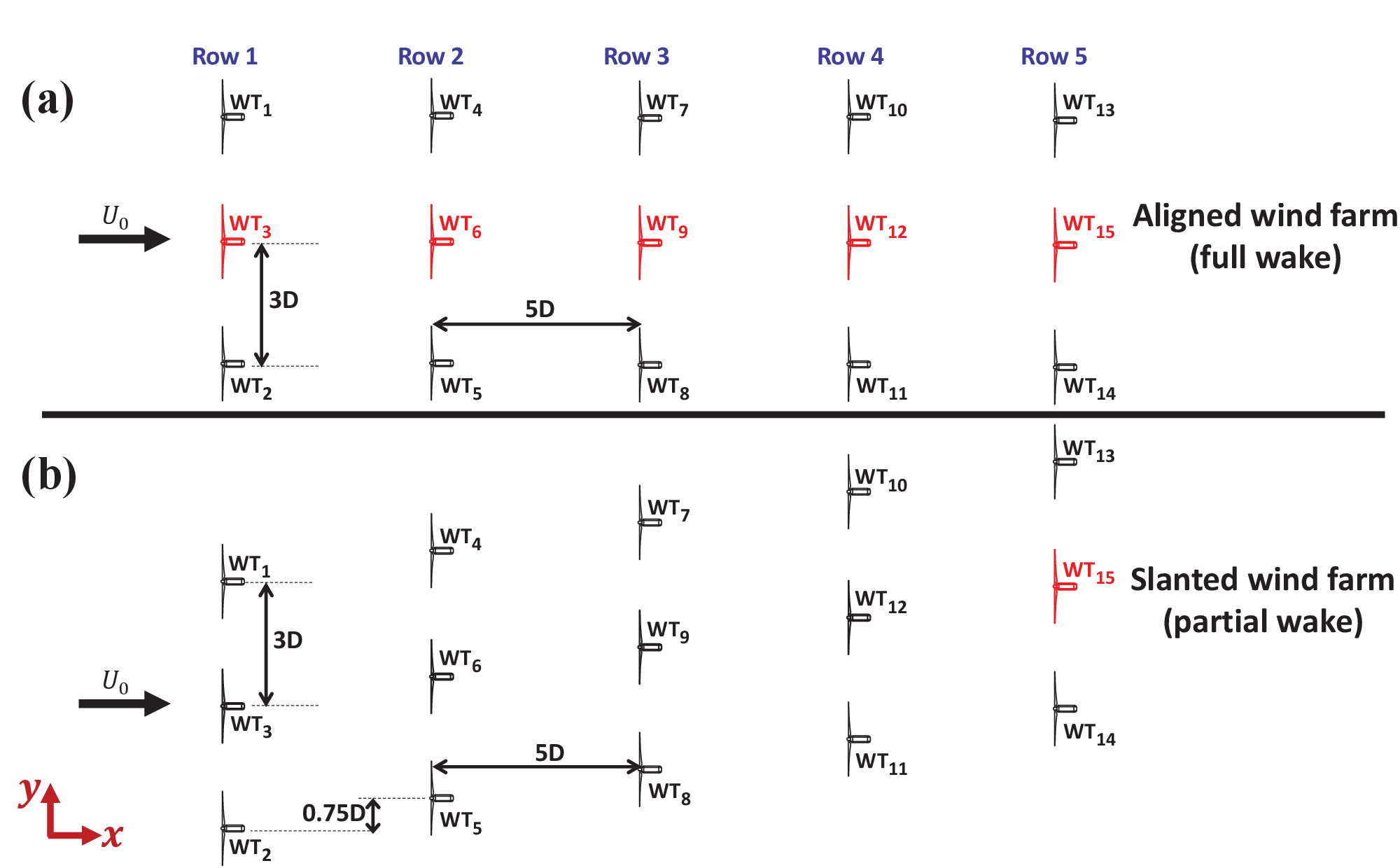}
		\caption{\color{black} Schematic top view of the simulated (a) aligned wind farm cases with the number of rows varying from one to five, and (b) the slanted wind farm with five rows. For each case, simulations were performed with and without the last middle turbine shown by the red colour (e.g., WT$_6$ for the $3\times2$ aligned wind farm or WT$_{15}$ for the $3\times 5$ slanted wind farm).}	
		\label{Fig:schematics_WFs}
	\end{center}
\end{figure}

We simulate cases of aligned wind farm array with streamwise inter-turbine spacing of $S_x=5D$ and spanwise inter-turbine spacing of $S_y=3D$, where $D$ is the rotor diameter.
The inter-turbine spacing is intentionally chosen to be small to create strong turbine interactions within wind farms and thereby make it easier to examine capabilities and limitations of the developed analytical model. All simulated aligned wind farms consist of three turbine columns, but the number of turbine rows in simulations varies from one to five. For each case, the simulations are performed with and without the turbine located in the last row and the middle column, shown by the red colour in figure \ref{Fig:schematics_WFs}a (e.g., WT$_6$ for  the $3\times 2$ array or WT$_{15}$ for the $3\times 5$ array).
Removing turbines from the domain allows us to systematically study the magnitude of different terms in (\ref{Eq:Nav-Stk_general-x5}).
We simulate another wind farm array with a slanted layout as shown in figure \ref{Fig:schematics_WFs}b.
This wind farm, hereafter called slanted wind farm, is chosen to study model predictions in partial wake conditions. The slanted wind farm was constructed from the aligned wind farm by laterally shifting each row by $0.75D$ with respect to the upstream row. Unlike the aligned wind farm, the simulations for the slanted wind farm are only performed for the $3\times 5$ array with and without WT$_{15}$ shown by the red colour in figure \ref{Fig:schematics_WFs}b.


\color{black}
\section{Derivation of wind farm analytical solution}\label{sec:model_derivation}
 
{\color{black}
The results from LES of the aligned wind farms are used to compute the integral quantities of  (\ref{Eq:Nav-Stk_general-x5}) in a cubic box surrounding WT$_n$. The width of the box ($y_b-y_a)$ is $500$ [m], which is wide enough to ensure that the wake of WT$_n$ is included. The vertical extent goes from $z_a=20$ [m]  to $z_b=300$ [m]. The streamwise extent of the integration box is from $x_a=x_n-2D$ to $x_b=x>x_n$. The first three terms on the right-hand side of (\ref{Eq:Nav-Stk_general-x5}) are surface integrals on a $y\!-\!z$ plane at $x=x_b$. The Reynolds shear stress term is a surface integral over horizontal planes at $z=z_b$ and $z_a$, while the mean flow shear term is the only volume integral in the equation.

We note that the filtered variables from LES are similar to those derived in  (\ref{Eq:Nav-Stk_general-x5}), however, there are some differences present, for example:
\begin{itemize}
    \item[-] the LES velocities are spatially filtered quantities,
    \item[-] the unresolved part of the Reynolds stresses is neglected,
    \item[-] the pressure includes the deviatoric part of the Reynolds stress tensor.
\end{itemize}

These differences \color{black} are expected to \color{black} cause a small difference in the momentum balance, which is captured in the residual term. \color{black}See \citet{sullivan1994,juneja1999,ghate2018} among others for a detailed discussion on the differences between pressure and velocity modelled by LES and their true values\color{black}. Figure \ref{Fig:integral_momentum} shows the integral terms in (\ref{Eq:Nav-Stk_general-x5}) computed from the LES of the aligned wind farm cases with different numbers of rows. For each case, the results are shown for the last turbine in the middle column shown by the red colour in figure \ref{Fig:schematics_WFs}a (e.g., WT$_6$ for Row 2 ($n=6$)). All terms are normalised by the value of the thrust force term. 

Figure \ref{Fig:integral_momentum} shows that the pressure term is the dominant term in the near wake, which is expected, as the sudden change in pressure at the rotor generates the turbine thrust force. 
The value of the momentum deficit term is clearly smaller than the thrust force in the near wake region. 
However, it increases with an increase of downstream distance and becomes comparable to the thrust force in the far wake region, especially for turbines in the first three rows. This is consistent with recent laboratory results reported by \citet{hulsman2020}. 
The next important term in the momentum equation is the Reynolds normal stress. This term is expected to be non-negligible for any wake flows, as also shown in prior studies of bluff-body wakes \citep{terra2017}. 
Next, we examine the two terms that are introduced by the incoming boundary layer. 
The value of the Reynolds shear stress term is relatively small for turbines in the first and second rows. It seems that its value slightly increases for the one in the last row, although it is still smaller than other dominant terms. Note that values of the terms, including $\overline{uw}_n$ in (\ref{Eq:Nav-Stk_general-x5a}) and $\overline{uw}_{n-1}$ in (\ref{Eq:Nav-Stk_general-x5b}), are expected to be significant due to the presence of the boundary layer \citep[][]{cal2010experimental,Bastankhah2017_Part_II}. 
However, our LES data suggest that the difference in their values does not seem to be large, at least for a finite-size wind farm with five rows of wind turbines.  
The mean flow shear term is the last term on the right-hand side of  (\ref{Eq:Nav-Stk_general-x5}). This term is the product of the vertical mean incoming flow shear and the vertical velocity component, mainly induced by the wake rotation. 
Figure \ref{Fig:integral_momentum} shows that this term is small but not negligible and its variation is relatively similar for turbines at different rows. We note that the effect of incoming wind veer is neglected in this analysis, given that its effect on the balance of momentum in the streamwise direction is expected to be insignificant. Indeed, veer becomes more important for the balance of momentum in the spanwise direction, which is not considered here. In the case of strong veer, the shape of the wake cross-section is skewed \citep{abkar2018veer}, which is out of the scope of the current study.

\begin{figure}
	\begin{center}
		\includegraphics[width=1\textwidth]{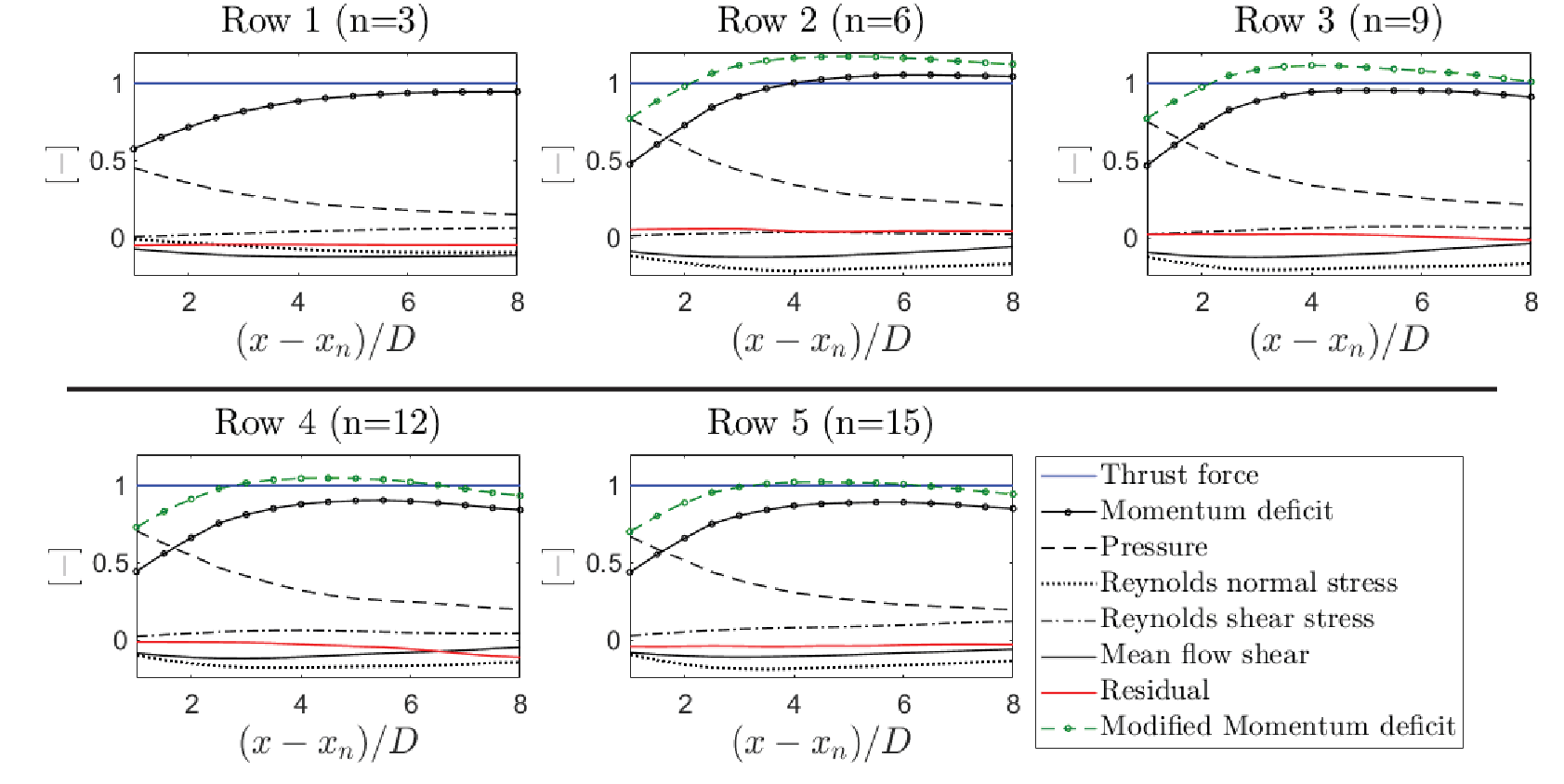}
		\caption{\color{black} Variation of different terms in (\ref{Eq:Nav-Stk_general-x5}), normalised by the thrust force term, for the last turbine in the middle column of aligned wind farms with different number of rows. The modified momentum deficit term is discussed later in \S\ref{sec:modified_momentum_deficit}.}	
		\label{Fig:integral_momentum}
	\end{center}
\end{figure}

It is important to note that for a very large wind farm that asymptotes to a so-called fully developed flow regime, changes in the streamwise direction can be almost neglected. In this case, the energy is mainly transported vertically from the top of the wind farm \citep{calaf2010large,abkar2013,meneveau2012}. 
However, even for this asymptotic case, the momentum deficit term in  (\ref{Eq:Nav-Stk_general-x5}) is far from being negligible as this term is the difference of the momentum deficit flux, with and without the turbine at the same downwind position, not the difference of the one before and after a turbine. 
Future work is indeed needed to systematically examine the significance of different terms in (\ref{Eq:Nav-Stk_general-x5}) (especially the momentum deficit, pressure and Reynolds shear stress terms) for very large wind farms.

}

\color{black}

\color{black}Based on the above discussion on the magnitude of different terms in (\ref{Eq:Nav-Stk_general-x5}), the below equation seems to be a reasonable approximation in the far wake of WT$_n$ (at least for moderate values of $n$):  
\begin{equation}\label{Eq:momentum_deficit_wf}
\rho\int U_{n}\left(U_0-U_{n}\right)\mathrm{d}A  - \rho\int U_{n-1}\left(U_0-U_{n-1}\right)\mathrm{d}A\approx T_n.
\end{equation}  
Equation (\ref{Eq:momentum_deficit_wf}) can be loosely named as \emph{conservation of momentum deficit}. \color{black}It states that the presence of a turbine induces a rise in the value of momentum deficit flux, and the magnitude of this rise is equal to the turbine thrust force. \color{black} It is important to bear in mind that the conservation of momentum deficit is not an intrinsic flow governing equation like those for mass and momentum that should always hold true. As shown earlier, (\ref{Eq:momentum_deficit_wf}) is just an approximate relation deduced from conservation of mass and momentum (\ref{Eq:Nav-Stk_general-x5}). Also note that (\ref{Eq:momentum_deficit_wf}) is not the same as the one used for single turbine wakes \citep[e.g.,][]{bastankhah2014new}. See \S\ref{sec:stw} for further discussion.

In the following, we aim to determine $U_n$ by solving (\ref{Eq:momentum_deficit_wf}). To achieve this goal, we need to use the assumption of self-similarity for velocity-deficit profiles in turbine wakes. \color{black} Unlike single isolated turbines, the definition of velocity deficit with respect to the incoming flow (i.e., $U_{in,n}$) is not suitable for turbines within wind farms. This definition can even lead to negative values of velocity deficit at the centre of the wake for very large downwind distances, where the wake velocity becomes greater than the incoming flow (\color{black} i.e., as $(x-x_n)\to\infty$, $U_n\to U_0\geq U_{in,n}$ \color{black}). Instead, we define the velocity deficit at a given position $\mathbf{X}=(x,y,z)$ downwind of WT$_n$ as the difference of the streamwise velocity in the absence and presence of WT$_n$ at $\mathbf{X}$; i.e.,
\begin{equation}\label{Eq:new_velocity deficit}
\Delta U_n(\mathbf{X})=U_{n-1}(\mathbf{X})-U_{n}(\mathbf{X}).
\end{equation}
\color{black}

\begin{figure}
	\begin{center}
		\includegraphics[width=1\textwidth]{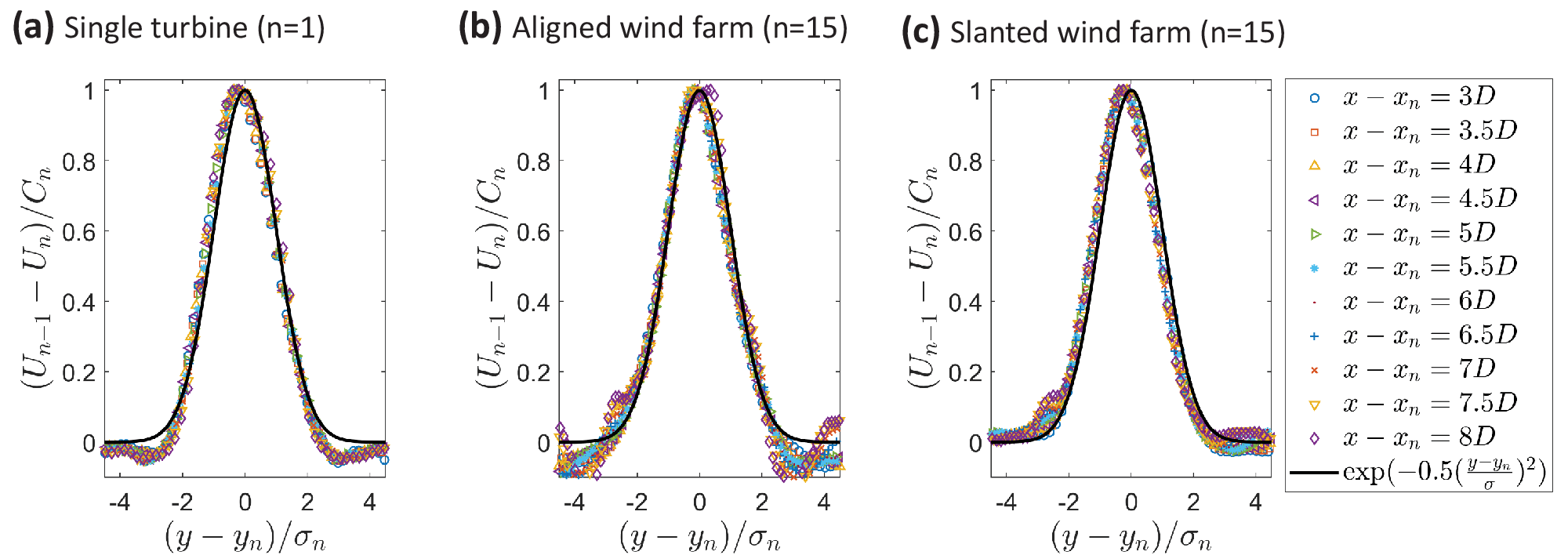}
		\caption{\color{black} Self-similar profiles of velocity deficit at different locations downwind of WT$_n$ for the LES data for (a) a single turbine ($n=1$), (b) the turbine in the fifth row (middle column) of the aligned wind farm ($n=15$) and (c) the turbine in the fifth row (middle column) of the slanted wind farm ($n=15$).}	
		\label{Fig:self_similarity2}
	\end{center}
\end{figure}

Figure \ref{Fig:self_similarity2} shows that with this definition of velocity deficit, the wake of a turbine in a wind farm exhibits a good degree of self-similarity, akin to a stand-alone turbine. As seen in the figure, velocity-deficit profiles collapse to a single curve for different downwind positions if the velocity deficit is normalised by the maximum velocity deficit $C_n$ and the distance from the wake centre is normalised by the characteristic wake half width $\sigma_n$. The results are shown in figure \ref{Fig:self_similarity2} for wakes of three turbines: (a) a stand-alone wind turbine, (b) the middle turbine in the last row of the aligned wind farm (i.e., $WT_{15}$), and (c) the middle turbine in the last row of the slanted wind farm (i.e., $WT_{15}$). It is also worth mentioning that a slight lateral wake deflection observed in the figure for the LES data is likely due to the interaction of rotating wake with the vertical inflow shear discussed by prior studies \citep[e.g.,][]{fleming2014evaluating,gebraad2014wind}. The developed analytical model in this paper does not take into account this slight wake deflection for zero-yawed turbines.
As $(U_{n-1}-U_{n})$ is self-similar, we can write
\begin{equation}\label{Eq:self-similarity2}
U_{n-1}-U_{n}=C_n(x)f_n\left(\frac{y}{\sigma_n(x)},\frac{z}{\sigma_n(x)}\right),
\end{equation}
where $f_n$ is the self-similar function. Shifting the index $n$ in (\ref{Eq:self-similarity2}) to $n-1$, $n-2$, ... $1$ leads to a set of equations as follows:
\begin{equation} \label{eq1}
\begin{split}
U_{n-2}-U_{n-1} & = C_{n-1}f_{n-1}, \\
U_{n-3}-U_{n-2} & = C_{n-2}f_{n-2}, \\
 & . \\
 & . \\
 & . \\
U_{0}-U_{1} & = C_{1}f_{1}. \\
\end{split}
\end{equation}
Adding (\ref{Eq:self-similarity2}) and (\ref{eq1}) results in
\begin{equation}\label{Eq:u_n-1}
U_{0}-U_{n}=\sum_{i=1}^{n}C_if_i.
\end{equation}
Using (\ref{Eq:self-similarity2}) and  (\ref{Eq:u_n-1}) to rearrange (\ref{Eq:momentum_deficit_wf}), we obtain
\begin{equation}\label{Eq:abc}
\bigintsss C_n f_n\left(U_{0}-C_n f_n-2\sum_{i=1}^{n-1}C_i f_i\right)\mathrm{d}A=\frac{T_n}{\rho}.
\end{equation}
To solve the above equation and find $C_n$, \color{black} a mathematical relation should be used to express $f_i$. \color{black} The boundary-free shear flow theory suggests a self-similar Gaussian solution for $f_i$ based on thin-shear simplification of RANS equations \citep{tennekes1972first}. \color{black} A Gaussian profile is shown in figure \ref{Fig:self_similarity2}a in comparison with the LES data \color{black} for a stand-alone wind turbine\color{black}. As seen in the figure, this can acceptably estimate self-similar velocity-deficit profiles for most of the wake, except for at the wake edges. A Gaussian profile is known to often overestimate the velocity deficit at wake boundaries \citep{bastankhah2014new,abkar2015influence,xie2014self,bastankhah2016experimental}. This velocity-deficit overprediction at wake edges can be explained by the fact that based on a Gaussian distribution, the velocity deficit decreases gradually by moving away laterally and vertically from the centre of the turbine, and it goes to zero at infinity. \color{black} However, the tangential vorticity shedding from the edge of the rotor induces a positive axial velocity in the outer region \citep{branlard2016,shapiro2019filtered, bontempo2019vorticity}, which leads to a slight flow speed-up, especially at short downwind distances\color{black}. This flow speed-up can cause lower than expected or even negative values of velocity deficit at the wake edges for a stand-alone turbine as shown in figure \ref{Fig:self_similarity2}a. \color{black} In a wind farm, a more pronounced flow speed-up may occur between adjacent turbine columns due to flow blockage effects. For instance, this is the case for the simulated aligned wind farm as shown in figure \ref{Fig:self_similarity2}b. On the other hand, our results suggest that \color{black} the flow speed-up for the last turbine in the slanted wind farm (figure \ref{Fig:self_similarity2}c) seems to be less significant. The magnitude of flow speed-up around wind turbines may depend on several factors such as turbine and inflow properties as well as wind farm layout geometries \citep{garrett2007efficiency,nishino2015}. The accurate estimation of flow speed-up is out of the scope of this study and so we use a Gaussian distribution for its simplicity. However, we acknowledge that this assumption can introduce errors in flow prediction at wake edges. See \S\ref{sec:model_predictions} for more discussion.  
\color{black}
With an assumption of the Gaussian distribution for the wake velocity deficit,
\begin{equation}\label{Eq:Gaussian2}
f_i=e^{-\frac{(y-y_i)^2}{2\sigma_i^2}}e^{-\frac{(z-z_i)^2}{2\sigma_i^2}}.
\end{equation}
Note that although the focus of this paper is turbines with no yaw angles, the developed model can be extended to cases with yawed turbines by replacing $y_i$ and $z_i$ in (\ref{Eq:Gaussian2}) with lateral and vertical positions of the wake centre at each streamwise position, respectively. 

Inserting (\ref{Eq:Gaussian2}) into (\ref{Eq:abc}), computing the surface integral with respect to $y$ and $z$ (from $-\infty$ to $+\infty$, neglecting ground effects) and \color{black} approximating $U_{0}$ with $U_h=U_0(z=z_h)$ \color{black} leads to
\begin{equation}\label{Eq:c_n}
C_n^2-2\left(U_{h}-\sum_{i=1}^{n-1}\lambda_{ni}C_i\right)C_n+\frac{T_n}{\rho \pi \sigma_n^2}=0,
\end{equation}
where $\lambda_{ni}$ is
\begin{equation}\label{Eq:lambda}
\lambda_{ni}=\frac{2\sigma_i^2 \mathrm{e}^{-\frac{\left(y_n-y_i\right)^2}{2\left(\sigma_n^2+\sigma_i^2\right)}}
\mathrm{e}^{-\frac{\left(z_n-z_i\right)^2}{2\left(\sigma_n^2+\sigma_i^2\right)}}
}{\sigma_n^2+\sigma_i^2}.
\end{equation}
\color{black} Note that for simplicity we assume in this work that the wake width is the same in both lateral and vertical directions. However, different values of lateral ($\sigma_{y}$) and vertical ($\sigma_{z}$) wake half widths can be used in (\ref{Eq:Gaussian2}). In this case, any $\sigma^2$ in (\ref{Eq:c_n}) and hereafter should be replaced with the product of  $\sigma_{y}$ and $\sigma_{z}$. \color{black}
 The quadratic (\ref{Eq:c_n}) has two solutions for $C_n$. The \color{black} physically \color{black} acceptable solution that decays with an increase of $\sigma_n$ is
\begin{equation}\label{Eq:final_eq}
\frac{C_n}{U_{h}}=\left(1-\sum_{i=1}^{n-1}\lambda_{ni}\frac{C_i}{U_{h}}\right)\left(1-\sqrt{1-\frac{c_{t,n}\left(\frac{<U_{n-1}>_{n,x_n}}{U_{h}}\right)^2}{8\left( \sigma_n/D \right)^2\left(1-\sum_{i=1}^{n-1}\lambda_{ni}\frac{C_i}{U_{h}}\right)^2}}\right),
\end{equation}
where $<>_{(i,x_j)}$ denotes spatial averaging over the frontal projected area of WT$_i$ at $x=x_j$, and the thrust coefficient of WT$_n$ is given by
\begin{equation}\label{Eq:T_n}
 c_{t, n}=\frac{8T_n}{\pi\rho D^2 <\!U_{n-1}\!\!>_{(n,x_n)}^2}.
\end{equation}
 Values of $C_1$, $C_2$, ... $C_n$ determined from (\ref{Eq:final_eq}) are inserted into (\ref{Eq:u_n-1}) to evaluate $U_n$. \color{black} While we use $<\!\!U_{n-1}\!\!>_{n,x_n}$ to relate $c_{t, n}$ to $T_n$ in (\ref{Eq:T_n}), one may approximate $<\!\!U_{n-1}\!\!>_{n,x_n}$ with $U_{n-1}(x_n,y_n,z_n)$ in (\ref{Eq:final_eq}) and (\ref{Eq:T_n}) for simplicity. \color{black} Note that ($\ref{Eq:final_eq}$) is a recursive sequence, so $C_n$ can be explicitly computed as a function of $C_i$ (for $i=1,$ ... $n-1$). To compute $C_1$ (i.e., $n=1$), \color{black} by setting $\lambda=0$, (\ref{Eq:final_eq}) is reduced to the one developed by \citet{bastankhah2014new} for a stand-alone turbine.


The dimensionless coefficient of $\lambda_{ni}$ in (\ref{Eq:final_eq}) has an interesting physical interpretation. It quantifies the contribution of WT$_i$ on the value of $C_n$ (i.e., the velocity deficit associated with WT$_n$). From (\ref{Eq:lambda}), $\lambda_{ni}$ depends on the wind farm layout and inflow conditions, and its value can vary from $0$ to $2$. As $|y_i-y_n|$ tends to infinity, $\lambda_{ni}$ tends to zero. This means that if the turbines are laterally distant from one another, they do not have any interaction, so (\ref{Eq:final_eq}) is reduced to the solution derived for a single turbine. As $y_i$ approaches $y_n$ (i.e, partial-wake conditions), the value of $\lambda_{ni}$ increases. Ultimately, at $y_i=y_n$ (i.e., full-wake conditions) and assuming $z_i=z_n$, $\lambda_{ni}$ becomes equal to $2\sigma_i^2/(\sigma_i^2+\sigma_n^2)$, which can take any value between one and two. In these conditions, $\lambda_{ni}$ tends to $2$ if $\sigma_i$ goes to infinity, which occurs when the turbine is immersed in a reasonably large upwind wake. On the other hand, the value of $\lambda_{ni}$ tends to unity if the upwind turbine wake has a size comparable to that of the turbine wake (i.e., $\sigma_i\approx \sigma_n$). The inflow properties also affect the value of $\lambda$. An increase in the level of atmospheric turbulence enhances the wake  \color{black} recovery \color{black} rate, which in turn leads to an increase in the value of $\lambda$. This occurs particularly for turbines in front rows. Therefore, the coefficient $\lambda$ obtained purely based on an analytical approach is analogous to empirical methods used in the literature to quantify the effects of upwind turbine wakes, such as finding the areas of wake overlap with each turbine. For instance, see the so-called `mosaic-tile' approach used by \citet{rathmann2006}, among others. 

The second exponential term in the right-hand side of (\ref{Eq:lambda}) is equal to unity for $z_i=z_n$. This is the common case in wind farms given that the turbines usually have the same hub height. We, however, leave this term in its original form because of (i) its potential use in imaging techniques to simulate ground effects \citep{Crespo1999} as well as (ii) its potential application in studying wind farms with variable hub heights (i.e., vertical staggering) \citep{Stevens2019verticla_staggering}. 

\color{black}It is also worth mentioning that in the case of $U_0=U_0(x,y,z)$, $U_h$ in (\ref{Eq:final_eq}) is substituted with $U_0(x_n,y_n,z_n)$. However, note that the developed solution is expected to provide acceptable estimations as long as values of $\textrm{d}U_0/\textrm{d}x$ and $\textrm{d}U_0/\textrm{d}y$ are not significant. Otherwise, the incoming flow heterogeneity induces non-negligible additional terms in (\ref{Eq:Nav-Stk_general-x5}), akin to the vertical mean flow shear term.

\color{black}
\section{Model predictions}\label{sec:model_predictions}
\color{black}
\begin{figure}
	\begin{center}
		\includegraphics[width=.9\textwidth]{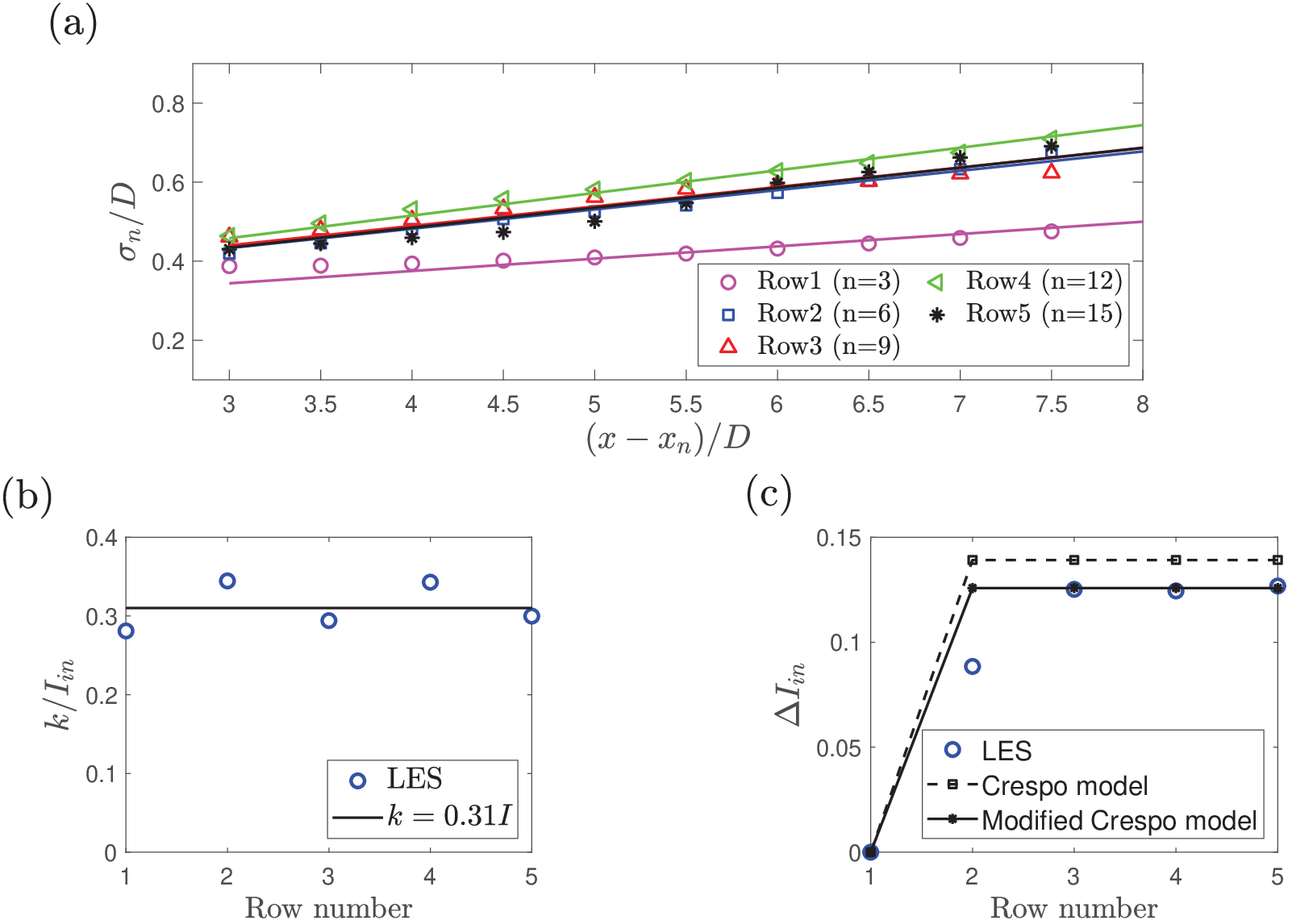}
		\caption{\color{black}(a) Variation of the normalised wake half width with the streamwise distance for the last turbine in the middle column of aligned wind farms with different number of rows. (b) The ratio of wake recovery rate $k$ to the incoming turbulence intensity $I_{in}$ for turbines at different rows based on the LES data. (c) Comparison of incoming added turbulence intensity predicted by the Crespo model \citep{crespo1996} and the suggested modified Crespo model with the LES data for turbines at different rows.}	
		\label{Fig:wake_recovery}
	\end{center}
\end{figure}
In this section, the developed model is used to predict the flow distribution in each of the two wind farms shown in figure \ref{Fig:schematics_WFs}, and the results are compared with the LES data. In order to compute predictions using the analytical model, an estimation of the wake recovery rate $k$ is required as the only empirical input of the model. Prior studies have suggested that the wake recovery rate for a single turbine is directly proportional to the turbulence intensity of the incoming flow \citep{Niayifar2016,carbajo2018,shapiro2019paradigm,zhan2020Iungo}. In the following, we examine the validity of this assumption for turbines of the aligned wind farm, for which LES were performed in the presence and absence of turbines at different rows.

The lateral velocity deficit profiles at different downstream distances of the middle turbine in each row are analysed using the definition of velocity deficit stated earlier as $\Delta U_n(\mathbf{X})=U_{n-1}(\mathbf{X})-U_{n}(\mathbf{X})$ for WT$_n$ (i.e., the difference between the flow with and without the presence of WT$_n$). A Gaussian distribution (\ref{Eq:Gaussian2}) is fitted to each velocity deficit profile in the horizontal plane at hub height in order to estimate the corresponding wake half width $\sigma_n$ for each downstream position. Figure \ref{Fig:wake_recovery}a shows the variation of wake half width with downwind distance for the aligned wind farms with different number of rows. Results show that the wake width is clearly smaller for the turbine in the first row, which is expected because of the lower level of turbulence intensity in the incoming flow. Most of the turbine wakes appear to have a linear expansion. The results for the third and fifth rows appear slightly less linear, which may be due to some uncertainty in the estimation of the wake width using a Gaussian curve fitting as discussed in prior studies \citep[e.g.,][]{quon2020nrel_wake_recovery}. However, a linear curve still provides satisfactory agreement. Subsequently, the value of $k$ for WT$_n$ is determined by fitting a linear curve with
\begin{equation}\label{eq:sigma_i}
   \sigma_n(x)/D=k\left(x-x_n\right)/D+\epsilon, 
\end{equation}  
where $\epsilon$ is the normalised initial wake half width given by $0.2\sqrt{\beta}$ and $\beta=\left(1+\sqrt{1-c_{t, n}})\right)/\left(2\sqrt{1-c_{t, n}}\right)$ \citep{bastankhah2014new}. Fitting the curve in this way ensures that the initial wake width is consistent throughout the model to give values of $k$ that are a fair representation of the wake recovery rate.

Following this, the proportionality of $k$ in relation to the incoming turbulence intensity $I_{in}$ is investigated. For $WT_n$, the value of $I_{in}$ is defined as the value of turbulence intensity at the rotor position in the absence of the rotor, i.e., $\sqrt{<\overline{uu}>_{(n-1,x_n)}}/<\!U\!>_{(0,x_n)}$. Figure \ref{Fig:wake_recovery}b shows that $k/I_{in}$ is not exactly identical for all turbines. However, the deviation of this ratio between different rows is not significant, and so the mean value of $k/I_{in}=0.31$ can be taken to approximate the relationship between the wake recovery rate $k$ and the incoming turbulence intensity $I_{in}$. This value is lower than those suggested in previous works, such as 0.38 in \citet{Niayifar2016}, 0.35 in \citet{carbajo2018} and 0.34 in \citet{zhan2020Iungo}. The discrepancy in the coefficient that relates $I_{in}$ to $k$ may be due to the wake recovery rate having some weak dependencies on incoming flow characteristics other than the incoming turbulence intensity (e.g., the integral length scale). It is not in the scope of this work to establish a complete relationship between the wake recovery rate and incoming turbulence intensity, as well as potentially other important factors. Nevertheless, this is an important topic and needs to be studied more rigorously in future research. This is further discussed in \S\ref{sec:summary}.

The final step in estimating the wake recovery rate for the analytical model requires an estimation of the incoming turbulence intensity for each turbine in a wind farm. The empirical relationship suggested by \citet{crespo1996} is used here for simplicity, although other, more detailed, empirical relations could be used instead, such as the one proposed by \citet{ishihara2018new}. The magnitude of incoming turbulence intensity $I_{in}$ for each turbine is taken as $\sqrt{{I_0}^2+\Delta I_{in}^2}$, where $I_0$ is the ambient turbulence intensity, and $\Delta I_{in}$ is the turbulence intensity added by upwind turbines . The value of $\Delta I_{in}$ for WT$_n$ due to the upwind turbine WT$_i$, where $i<n$, is estimated with the relationship proposed by \citet{crespo1996} (hereafter referred to as the Crespo model): $\Delta I_{in}=0.73a_i^{0.83}I_0^{0.03}\left[(x_n-x_i)/D\right]^{-0.32}$, where $a_i$ is the induction factor of WT$_i$ given by $0.5(1-\sqrt{1-c_{t,i}})$. This relationship is used in conjunction with the geometric method suggested by \citet{Niayifar2016}. This involves finding the fraction of overlap area between the turbine rotor and upstream wakes to determine the added turbulence intensity due to each upstream turbine, followed by taking the maximum value of $\Delta I_{in}$ (i.e., the upstream turbine with the largest impact on the incoming flow). By using this method, predictions show that turbines in the second row and subsequent rows experience the same incoming turbulence intensity. However, this is not completely valid for this wind farm case as the LES data show that the value of incoming turbulence intensity for the second row is less than that of subsequent rows. Additionally, this method slightly overestimates $I_{in}$ for turbines deep inside a wind farm. In order to account for this offset, the Crespo model is employed for the analytical model with a slightly modified constant term to obtain the relationship $\Delta I_{in}=0.66a_i^{0.83}I_0^{0.03}\left[(x_n-x_i)/D\right]^{-0.32}$. As demonstrated in Figure \ref{Fig:wake_recovery}c, this modified version gives closer predictions of $I_{in}$ relative to the LES data.

 \begin{figure}
	\begin{center}
		\includegraphics[width=1\textwidth]{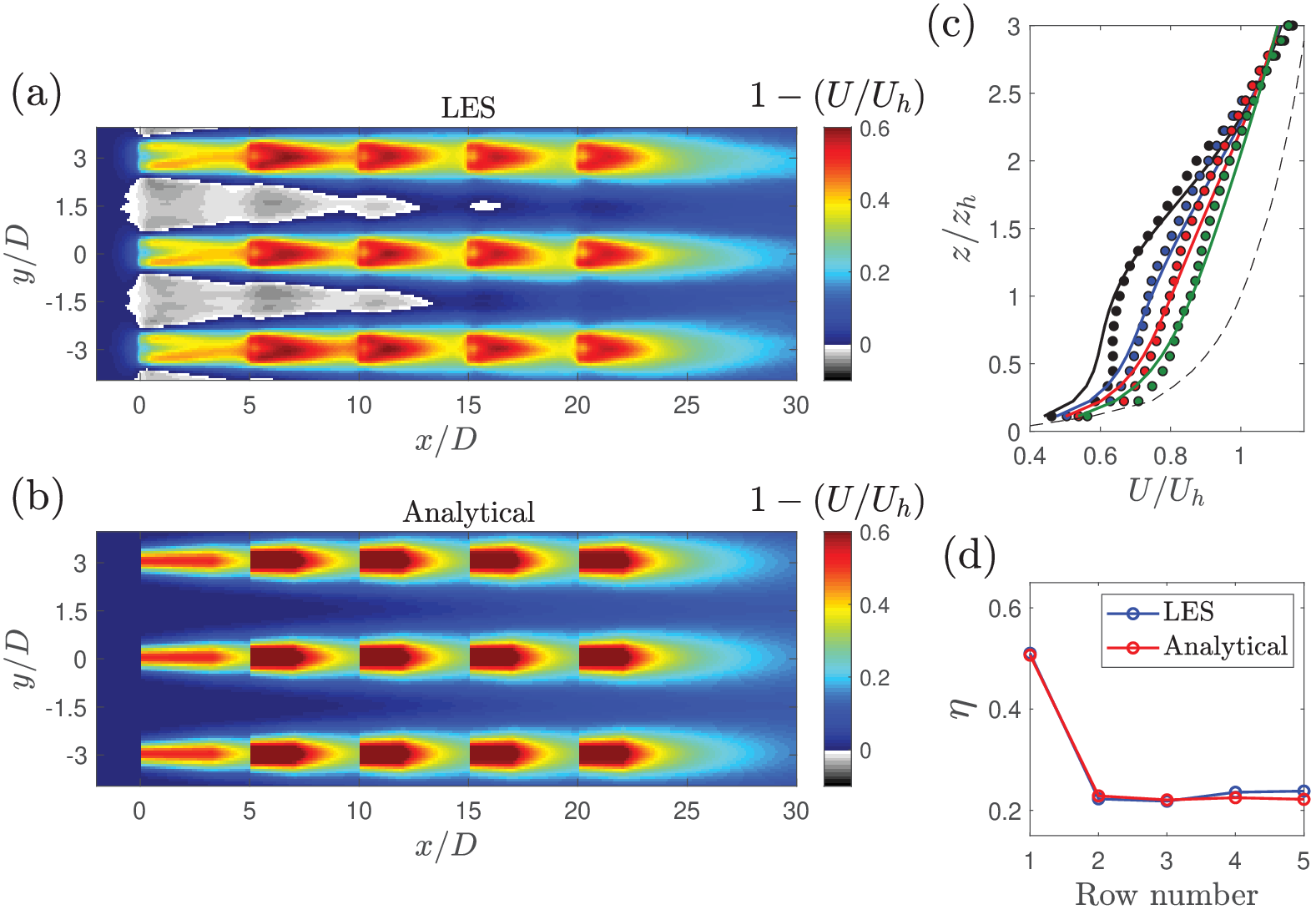}
		\caption{\color{black} Contours of $(U-U_h)$, normalised by the incoming velocity at hub height $U_h$, at a horizontal plane at hub height for the aligned wind farm (full-wake case) based on (a) LES and (b) analytical model predictions. (c) Vertical profiles of the normalised streamwise velocity predicted by the LES (circles), and the analytical model (solid lines) at various distances downwind of WT$_{15}$: $x-x_{15}=$ 4D (black), 6D (blue), 8D (red), and 12D (green). The dashed line shows the incoming velocity profile.  (d) Efficiency $\eta$ of turbines in the middle column, where $\eta$ for each turbine is defined as $P/(0.5\rho A U_h^3)$.}	
		\label{Fig:fw_wind_farm}
	\end{center}
\end{figure} 

Next, the wind farm flow distribution is estimated using (\ref{Eq:u_n-1}), (\ref{Eq:lambda}) and (\ref{Eq:final_eq}). Note that \color{black} the model can only provide reliable predictions in the far-wake region of wind turbines, where velocity-deficit profiles are self-similar, and also the assumptions made to develop the approximate relation of (\ref{Eq:momentum_deficit_wf}) are acceptable. At very short downwind distances, the term under the square root in (\ref{Eq:final_eq}) becomes negative and, as a result of this, (\ref{Eq:final_eq}) provides complex values. Prediction of the flow in the near wake region is not the objective of this work because it is unlikely for a turbine to be in the near wake of another turbine in realistic situations. However, the complex output of (\ref{Eq:final_eq}) might pose a practical issue in implementing the model to determine the incoming velocity for downwind turbines in some cases, \color{black} where two turbines are laterally far from each other but have similar streamwise locations\color{black}. In order to address this issue, any value of $C_n$ predicted by (\ref{Eq:final_eq}), which is either complex or bigger than $C_{n0}$ is replaced by $C_{n0}$, where $C_{n0}$ is the maximum theoretical velocity deficit based on the Betz theory and is given by $2a_n\!\!<\!U_{n-1}\!\!>_{(n, x_n)}$  \citep{manwell2010wind}. We adopt this approach for its simplicity, but if the goal is to realistically capture the near-wake region, recent models in the literature \citep[e.g.,][among others]{shapiro2019paradigm, double_gaussian2020, blondel2020} can be consulted. 

\color{black}

Finally, we discuss model predictions against the LES data. Figures \ref{Fig:fw_wind_farm}a and b show contours of ($U_h-U$), normalised by the inflow velocity at hub height $U_h$ for the $3\times 5$ aligned wind farm (i.e., the full-wake case). A different colourscale is used for negative values of ($U_h-U$) to indicate the speed-up region between adjacent turbine columns. This speed-up region appears to be evident between the primary rows of the wind farm, diminishing after roughly three rows of wind turbines. As mentioned earlier in \S\ref{sec:model_derivation}, this speed-up region is not accounted for in the developed model due to the assumption that the wake velocity deficit profile is Gaussian. For the aligned wind farm, the speed-up region does not largely interact with downwind turbines, so its impact is expected to be insignificant. However, this is not always the case, as discussed later for the case of the slanted wind farm.

Overall, figure \ref{Fig:fw_wind_farm}b shows that the model predictions of velocity in the far-wake region are in good agreement with the LES data. This is also confirmed in figure \ref{Fig:fw_wind_farm}c showing vertical profiles of normalised streamwise velocity at different positions downwind of the middle turbine in the last row (i.e., WT$_{15}$). Although the results show good agreement for the far-wake region, the figure shows that the model slightly underpredicts the velocity in the lower half of the wake, which becomes more apparent at short downwind distances (e.g., at $x-x_{15}=4D$). This discrepancy may be due to the uncertainty in the estimation of the wake recovery rate, as discussed earlier, or due to other terms being neglected in the right-hand side of the momentum equation (\ref{Eq:Nav-Stk_general-x5}). For cases where the momentum deficit term is less than the thrust force, assuming equality between these two terms will result in an overestimate of velocity deficit, which occurs particularly at short downwind distances. This is further discussed in \S\ref{sec:modified_momentum_deficit}. Additionally, it was assumed for simplicity that the wake recovery rate is the same in both the lateral and vertical directions. However, in the vertical direction, the wake flow may be affected by the mean shear of the incoming flow and presence of the ground resulting in a different value of $k$, as suggested by prior studies \citep{abkar2015influence,xie2014self}.

Finally, figure \ref{Fig:fw_wind_farm}d shows the efficiency of each middle turbine at different rows for both the LES data and predictions from the analytical solution. The efficiency $\eta$ of WT$_n$ is defined as $P_n/(0.5\rho U_h^3 A )$, where $P_n$ is the power extracted by WT$_n$ and $A$ is the area swept by the turbine blades. The turbine efficiency can be rewritten as $C_p<\!U\!>_{(n-1,x_n)}^3\!/U^3_h$, where $C_p$ is the power coefficient of the turbine. For analytical model predictions, the value of $<\!U\!>_{(n-1,x_n)}$ is estimated by the model, whereas $C_p$ for each turbine is estimated from figure \ref{Fig:ct_cp}. Overall, there is a good agreement between the efficiency predicted by the analytical model and the LES data, despite a slight underprediction of the efficiency for turbines in the last two rows. One possible explanation for this underprediction is the assumption of equality in (\ref{Eq:momentum_deficit_wf}), as discussed previously. As shown in figure \ref{Fig:integral_momentum}, this assumption is less accurate for turbines at the end of the wind farm, compared to those in primary rows.  

 \begin{figure}
	\begin{center}
		\includegraphics[width=1\textwidth]{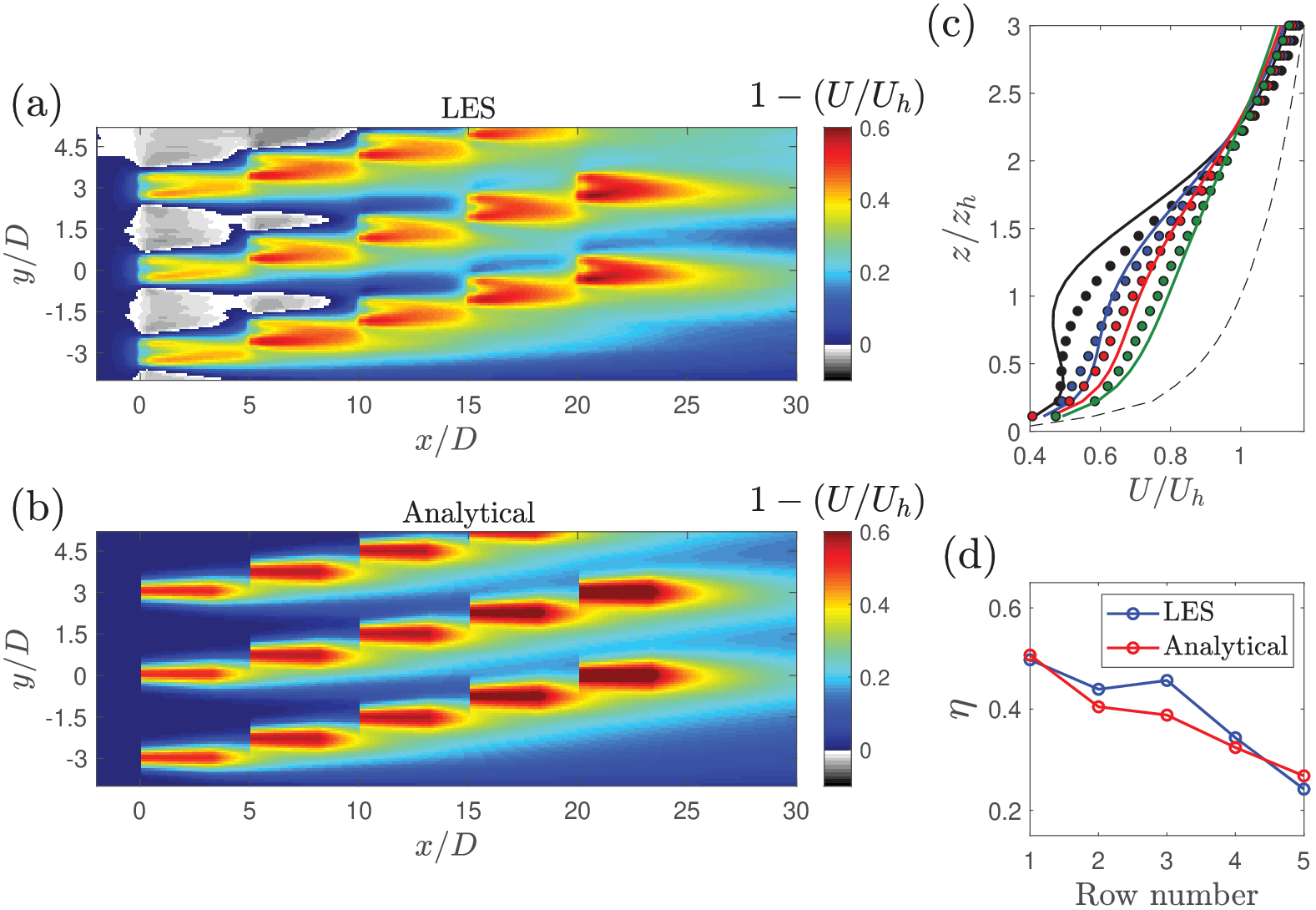}
		\caption{\color{black} Contours of $(U-U_h)$, normalised by the incoming velocity at hub height $U_h$, at a horizontal plane at hub height for the slanted wind farm (partial-wake case) based on (a) LES and (b) analytical model predictions. (c) Vertical profiles of the normalised streamwise velocity predicted by the LES (circles), and the analytical model (solid lines) at various distances downwind of WT$_{15}$: $x-x_{15}=$ 4D (black), 6D (blue), 8D (red), and 12D (green). The dashed line shows the incoming velocity profile. (d) Efficiency $\eta$ of turbines in the middle column, where $\eta$ for each turbine is defined as $P/(0.5\rho A U_h^3)$.}	
		\label{Fig:pw_wind_farm}
	\end{center}
\end{figure}

For the case of the slanted wind farm, figures \ref{Fig:pw_wind_farm}a and b show contours of ($U_h-U$), normalised by $U_h$, in the horizontal plane at hub height for LES data and the analytical model predictions, respectively. From figure \ref{Fig:pw_wind_farm}c, model predictions show fairly good agreement with the LES data in the far wake of the turbine in the last row (i.e., WT$_{15}$), but it underpredicts the velocity at short downwind distances. Figure \ref{Fig:pw_wind_farm}d shows the comparison between predictions of turbine efficiency from the analytical model and the LES. Model predictions for turbines in rows 4 and 5 show more satisfactory agreement than those for rows 2 and 3. The observed discrepancy in efficiency for rows 2 and 3 are most likely due to the flow speed-up demonstrated in figure \ref{Fig:pw_wind_farm}a. As discussed in \S\ref{sec:model_derivation}, the analytical model does not capture flow speed-up between adjacent turbine columns due to the assumption of Gaussian distribution for velocity-deficit profiles. This leads to underprediction of efficiency for turbines in the second and third rows, as shown in figure \ref{Fig:pw_wind_farm}d. The turbines in the last two rows do not seem to considerably benefit from the flow speed-up because they are in wakes of adjacent column. A similar observation is made in figure \ref{Fig:self_similarity2}c showing self-similar profiles for the turbine in the last row. 

These results highlight the limitation of this model (and essentially most other existing engineering wake models in the literature) for estimating the incoming flow of turbines that are subject to the speed-up between neighbouring columns. However, our LES data suggest that this speed-up effect appears to be important mostly at primary rows of wind farms and becomes less significant deep inside a wind farm. Moreover, the speed-up effect is expected to have less of an impact for wind farms with larger inter-turbine spacing. An important area of future research would be to evaluate the impact of flow speed-up on the power production for wind farms with various inflow conditions and layout configurations.

\color{black}

\color{black}

\color{black}
\section{A modified version of the conservation of momentum deficit}\label{sec:modified_momentum_deficit}
As shown in figure \ref{Fig:integral_momentum}, the magnitude of the momentum deficit term in (\ref{Eq:Nav-Stk_general-x5}) is clearly less than the thrust force in the near wake for all turbines. In the far wake, its value is closer to the turbine's thrust force. However, for turbines deep inside the wind farm (i.e., rows 4 and 5), the momentum deficit term is still slightly smaller than the thrust force even in the far wake region. This section attemps to modify the momentum deficit term such that this term's value would be more comparable to the thrust force magnitude over a broader range of streamwise distances and turbine rows. We start with rewriting the left-hand side of (\ref{Eq:momentum_deficit_wf}), so we obtain
\begin{multline}\label{eq:deficit_two_ways}
    \int U_{n}\left(U_0-U_{n}\right)\mathrm{d}A  -\int U_{n-1}\left(U_0-U_{n-1}\right)\mathrm{d}A=\\\int U_n\left(U_{n-1}-U_n\right)\textrm{d}A-\int\left(U_{n-1}-U_{n}\right)\left(U_0-U_{n-1}\right)\textrm{d}A.
\end{multline}
The second term on the right-hand side of (\ref{eq:deficit_two_ways}) is negative as $U_n<U_{n-1}$ and $U_{n-1}\leq U_0$. Therefore, one can write
\begin{equation}\label{eq:deficit_two_ways_v2}
    \underbrace{\left[\int U_{n}\left(U_0-U_{n}\right)\mathrm{d}A  -\int U_{n-1}\left(U_0-U_{n-1}\right)\mathrm{d}A\right]}_\text{Original Momentum Deficit Term}\leq\underbrace{\int U_n\left(U_{n-1}-U_n\right)\textrm{d}A.}_\text{Modified Momentum Deficit Term}
\end{equation}
The equality in (\ref{eq:deficit_two_ways_v2}) holds everywhere for $n=1$. For $n>1$, the equality holds only at large downwind distances where $U_n\to U_{n-1}\to U_0$, while the difference between the two sides of inequality is larger at shorter downwind distances. Figure \ref{Fig:integral_momentum} shows the variation of the modified momentum deficit term with the streamwise distance for turbines at different rows. While the value of this modified term is bigger than the thrust force in the far wake of turbines in primary rows (i.e., rows 2 and 3), its value in the far wake of those in rows 4 and 5 is closer to the thrust force than the original relation. Moreover, the modified term is closer to the thrust force at short downwind distances for all turbines. Therefore, we introduce a modified version of the conservation of momentum deficit as follows:
\begin{equation}\label{eq:modified_momentum deficit}
    \rho\int U_n\left(U_{n-1}-U_n\right)\textrm{d}A\approx T_n.
\end{equation}
Based on the LES data presented in figure \ref{Fig:integral_momentum}, this equation is expected to better hold the equality between the thrust and momentum deficit term at short downwind distances for all turbines and at long downwind distances for those deep inside a wind farm. 

\begin{figure}
	\begin{center}
		\includegraphics[width=.75\textwidth]{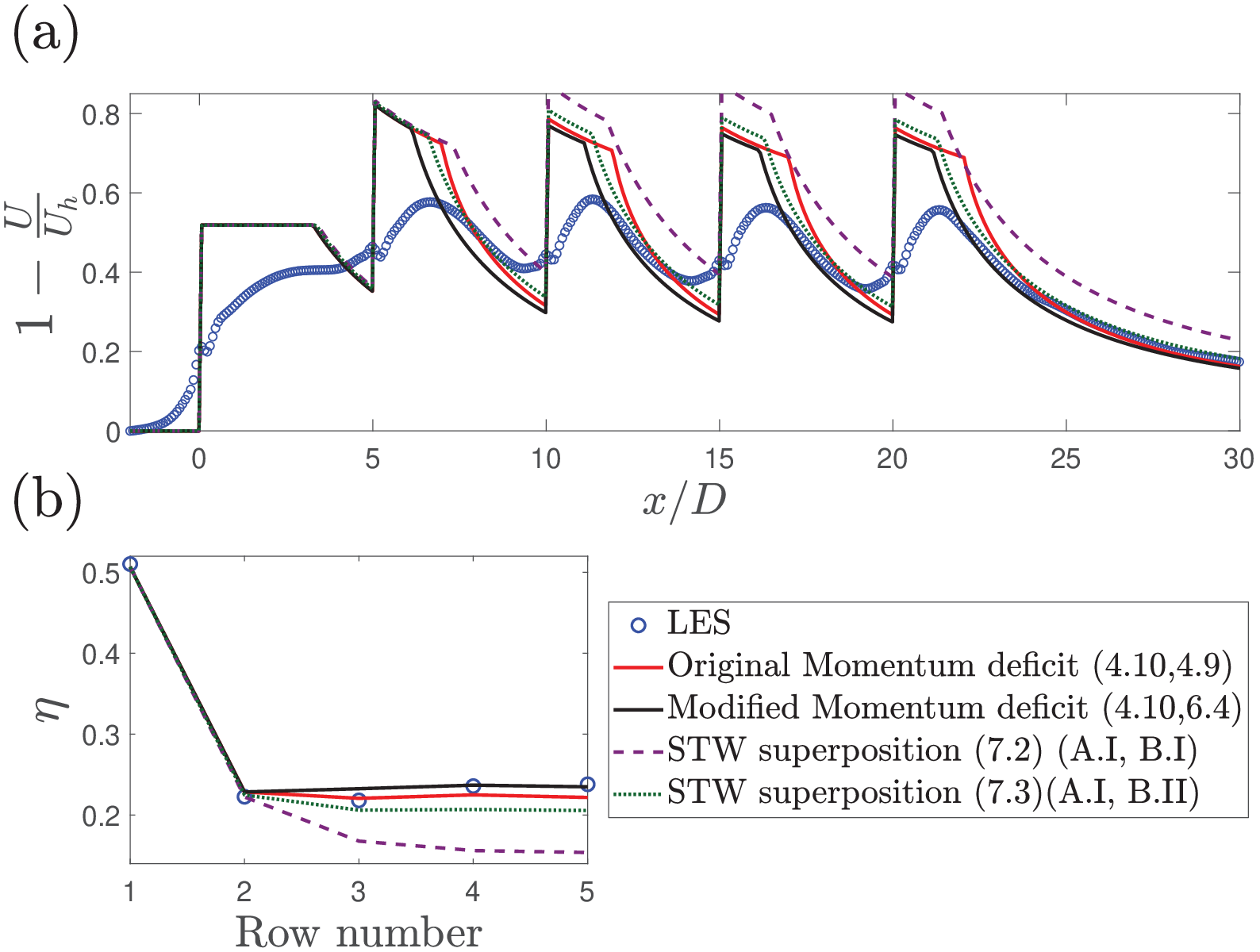}
		\caption{\color{black} (a) Variation of the centreline $(U_h-U)/U_h$ for the middle column of the $3\times 5$ aligned wind farm. Turbines are located at $x/D=0,5,10,15$ and $20$. (b) Efficiency $\eta$ of turbines in the middle column, where $\eta$ for each turbine is defined as $P/(0.5\rho A U_h^3)$. More details about STW superposition methods shown in the figure can be found in table \ref{Table}.}	
		\label{Fig:stw_comparison}
	\end{center}
\end{figure}

Equation (\ref{eq:modified_momentum deficit}) can be solved in the same way that we earlier solved (\ref{Eq:momentum_deficit_wf}) in \S\ref{sec:model_derivation} to find $U_n$. By doing so, we find that the solution of (\ref{eq:modified_momentum deficit}) is the same as (\ref{Eq:final_eq}). The only difference with the original solution is that $\lambda_{ni}$, based on the modified version, is half of the one obtained for the original equation. Therefore, using (\ref{Eq:final_eq}) in conjunction with (\ref{Eq:u_n-1}) and the modified definition of
\begin{equation}\label{Eq:lambda_modified}
\lambda^{\textrm{modified}}_{ni}=\frac{\sigma_i^2 \mathrm{e}^{-\frac{\left(y_n-y_i\right)^2}{2\left(\sigma_n^2+\sigma_i^2\right)}}
\mathrm{e}^{-\frac{\left(z_n-z_i\right)^2}{2\left(\sigma_n^2+\sigma_i^2\right)}}
}{\sigma_n^2+\sigma_i^2},
\end{equation}
forms the solution of the modified version of conservation of momentum deficit (\ref{eq:modified_momentum deficit}). Figure \ref{Fig:stw_comparison}a shows the variation of the centreline $(U_h-U)/U_h$ for turbines in the middle column of the aligned wind farm, while figure \ref{Fig:stw_comparison}b shows the efficiency of these turbines. It is worth remembering that model predictions in the upwind induction region and immediately behind the turbines are not valid. The figure shows that far-wake predictions, based on the original and modified versions, do not largely deviate from each other (e.g., compare flow predictions downwind of the last turbine at $(x-x_{15}>5D$). This is a good sign showing that model predictions are not highly sensitive to inaccuracies arising from using the approximate form of (\ref{Eq:Nav-Stk_general-x5}). However, a closer inspection of figure \ref{Fig:stw_comparison} confirms what we inferred earlier based on the results of figure \ref{Fig:integral_momentum}. The original momentum deficit works better in the far wake of turbines at primary rows (e.g., see $\eta$ for row 3 in figure \ref{Fig:stw_comparison}b). On the other hand, the modified version provides better predictions at short downwind distances for all turbines. It can also better predict the efficiency of turbines deep inside a wind farm (e.g., see $\eta$ for row 5 in figure \ref{Fig:stw_comparison}b). This comparison suggests that, overall, the modified version of the conservation of momentum deficit might be a slightly better choice, at least for turbines deep inside a wind farm. Future studies can perform similar analyses using more numerical and experimental data for larger wind farms to examine the accuracy of this conclusion.
\color{black}
\section{Superposition of single turbine wake (STW) models}\label{sec:stw}
In this section, we examine the validity of wake superposition techniques commonly used in the literature. To achieve this goal, we discuss approximations and assumptions that need to be made in order to derive superposition of STW models from \color{black} the solution of conservation of momentum deficit derived in \S\ref{sec:model_derivation} for a wind turbine in a wind farm\color{black}.

As discussed in \S\ref{sec:intro}, the common approach in existing superposition techniques is to treat each turbine separately and find the value of its wake velocity deficit at a given location. The values of velocity deficit caused by all turbines at that given location are then combined (either linearly or non-linearly) to find the total velocity deficit. By doing this, it is inadvertently assumed that the wind velocity downwind of a turbine subtracted from the one at the same location in the absence of the turbine can be expressed in a form similar to (\ref{Eq:self-similarity2}). As shown in figure \ref{Fig:self_similarity2}, this is a valid assumption for a turbine within a wind farm. 
\color{black}Next, we try to develop the STW model from the modified version of the conservation of momentum deficit (\ref{eq:modified_momentum deficit}). \color{black} If we approximate $U_{n-1}$ in this equation with $<\!U_{n-1}\!\!>_{(n,x)}$ at each $x$, (\ref{Eq:self-similarity2}), (\ref{Eq:Gaussian2}) and (\ref{Eq:T_n}) can be used to obtain
\begin{equation}\label{Eq:cn_stw_v1}
    C_n=<\!U_{n-1}\!\!>_{(n,x)}\left(1-\sqrt{1-\frac{c_{t,n}}{8(\sigma_n/D)^2}\frac{<\!U_{n-1}\!>^2_{(n,x_n)}}{<\!U_{n-1}\!>^2_{(n,x)}}}\right).
\end{equation}  

Next, we neglect the velocity ratio under the square root on the right-hand side of (\ref{Eq:cn_stw_v1}). Moreover, $<\!U_{n-1}\!\!>_{(n,x)}$ (i.e., the first term on the right-hand side of (\ref{Eq:cn_stw_v1})) is substituted with $U_{h}$. This leads to
\begin{equation}\label{Eq:cn_stw_BI}
    C_n=U_{h}\left(1-\sqrt{1-\frac{c_{t,n}}{8(\sigma_n/D)^2}}\right),
\end{equation}
which is the original STW model proposed by \citet{bastankhah2014new}. The velocity ratio of  $\left(<\!U_{n-1}\!>_{(n,x_n)}\!\!/\!<\!U_{n-1}\!>_{(n,x)}\right)$ tends to unity only as $(x-x_n)\rightarrow 0$ and is less than one at positive values of $(x-x_n)$ in most cases. On the other hand, the velocity ratio of $\left(<\!U_{n-1}\!>_{(n,x)}\!\!/U_{h}\right)$ tends to unity only as $(x-x_n)\rightarrow \infty$ and is less than one at definite values of $(x-x_n)$. Therefore, the two approximations made to develop  (\ref{Eq:cn_stw_BI}) from (\ref{Eq:cn_stw_v1}) lead to an overprediction of velocity deficit. \color{black} Flow predictions based on (\ref{Eq:cn_stw_BI}) and the linear superposition method (i.e., A.I in table \ref{Table}) are shown in figure \ref{Fig:stw_comparison}. One can observe that using this STW superposition model clearly leads to overpredictions of velocity deficit (see figure \ref{Fig:stw_comparison}a) and, consequently, underprediction of turbine efficiency (see figure \ref{Fig:stw_comparison}b). Similar observations were made by prior studies \citep[e.g.,][]{Niayifar2016,zong2020}.\color{black}

 In (\ref{Eq:cn_stw_BI}), $C_n$ is proportional to $U_{0}$, which is equivalent to the superposition method B.I shown in table \ref{Table}. To develop the method B.II in table \ref{Table}, we follow the same approach to derive (\ref{Eq:cn_stw_BI}) from (\ref{Eq:cn_stw_v1}), except $<\!U_{n-1}\!\!>_{(n,x)}$ in (\ref{Eq:cn_stw_v1}) is now replaced with $<\!U_{n-1}\!\!>_{(n,x_n)}$ (i.e., local incoming velocity), so we obtain
\begin{equation}\label{Eq:cn_stw_BII}
    C_n=<\!U_{n-1}\!\!>_{(n,x_n)}\left(1-\sqrt{1-\frac{c_{t,n}}{8(\sigma_n/D)^2}}\right).
\end{equation}
As often $\left(<\!\!U_{n-1}\!\!\!>_{(n,x_n)}\right)\leq\left(<\!\!U_{n-1}\!\!>_{(n,x)}\right)\leq \!\!U_h$ for $x\!\!\geq\!\!x_n$, the magnitude of velocity-deficit overprediction from (\ref{Eq:cn_stw_BII}) is significantly less than that from  (\ref{Eq:cn_stw_BI}), which is also seen in figure \ref{Fig:stw_comparison}.  

In comparison with the solution of the original conservation of momentum deficit (\ref{Eq:final_eq}), (\ref{Eq:cn_stw_BII}) underpredicts the velocity deficit at short downwind distances and overpredicts it at large ones, but overall its predictions do not largely deviate from those of (\ref{Eq:final_eq}). However, it is important to note that this is not because (\ref{Eq:cn_stw_BII}) is based on flow physics for the wake of a turbine within a wind farm. Several approximations made to develop (\ref{Eq:cn_stw_BII}) from the original solution have opposing effects on the prediction of wake velocity deficit, so their effects are cancelled out by each other to some extent. \color{black}Figure \ref{Fig:stw_comparison}b shows that using (\ref{Eq:cn_stw_BII}) in conjunction with the linear superposition method induces an error of about $3\%-4\%$ in predictions of turbine efficiency for the studied wind farm. \citet{zong2020} stated that the error of this superposition model might be more significant for larger wind farms. \color{black} Therefore, the use of the analytical wind farm solution (\ref{Eq:final_eq}) is recommended because it is not computationally more expensive than empirical STW superposition models. Nevertheless, if one tends to use superposition of STW models, (\ref{Eq:cn_stw_BII}) certainly provides more acceptable estimations than (\ref{Eq:cn_stw_BI}).

\begin{figure}
	\begin{center}
		\includegraphics[width=.9\textwidth]{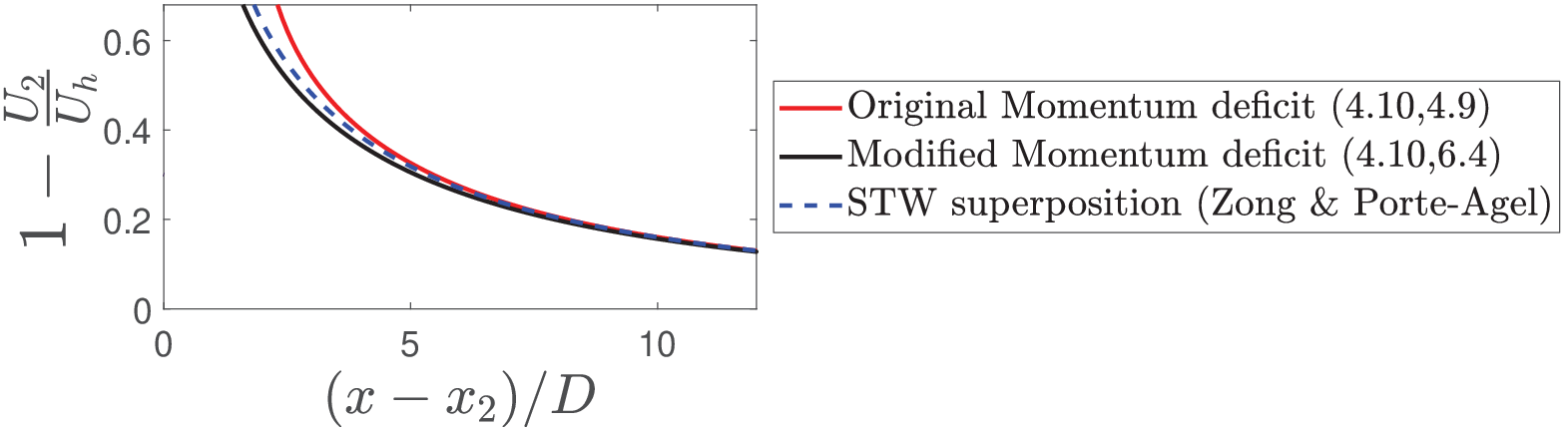}
		\caption{\color{black} Variation of centreline $(U_h-U_2)/U_h$ downwind of a turbine operating in full wake of another turbine.}	
		\label{Fig:stw_zong}
	\end{center}
\end{figure} 

Another important aspect of superposition techniques is the method used to superpose the velocity deficit caused by each turbine. Bearing in mind (\ref{Eq:u_n-1}), it is evident that the \emph{linear} superposition method (i.e., A.I in table \ref{Table}) is in better agreement with the analytical solution. However, the \emph{root-sum-square} method (A.II in table \ref{Table}) is more likely to provide better predictions if the velocity deficit caused by each single turbine is overestimated. Overestimation of velocity deficit for turbine wakes in wind farms mainly occurs for two reasons: (i) using the global incoming velocity $U_{h}$ as the reference velocity to compute the velocity deficit for each turbine (\ref{Eq:cn_stw_BI}), and (ii) estimating the wake recovery rate based only on inflow atmospheric conditions. The latter is expected to underestimate the wake recovery rate and thereby overpredict the velocity deficit, because it does not take into account a faster wake recovery due to the turbulence added by upwind turbines. This may explain why the \emph{root-sum-square} method has been customarily popular in the literature as either one or both of the above-mentioned assumptions can be commonly found in prior wake modelling studies \citep[e.g,][among others]{Katic1987}.

\color{black} Finally, we discuss the recent model developed by \citet{zong2020}. This model is based more on wake flow physics than other common superposition methods discussed earlier. They developed an iterative method to satisfy the following two equations:
\begin{gather} 
\rho\int U_{i}\left(U_{in,i}-U_{i}\right)\mathrm{d}A=T_i,\;\;\;\;\;i=\{1,2,...\;n\},\label{eq:zong1}\\
\rho\int U_{n}\left(U_0-U_{n}\right)\mathrm{d}A=\sum_{i=1}^n T_i.\label{eq:zong2}
\end{gather}
While (\ref{eq:zong2}) satisfies the conservation of momentum deficit for the whole wind farm, strictly speaking, (\ref{eq:zong1}) is not equal to the conservation of momentum deficit (\ref{Eq:momentum_deficit_wf}) for a turbine within a wind farm. In fact, one can show that
\begin{multline}\label{eq:zong3}
\small
     \underbrace{\int U_{i}\left(U_0-U_{i}\right)\mathrm{d}A  -\int U_{i-1}\left(U_0-U_{i-1}\right)\mathrm{d}A}_\text{Original Momentum Deficit}\approx \int U_{i}\left(U_0-U_{i}\right)\mathrm{d}A  -\int U_{in,i}\left(U_0-U_{in,i}\right)\mathrm{d}A=\\\int U_i\left(U_{in,i}-U_i\right)\textrm{d}A-\int\left(U_{in,i}-U_{i}\right)\left(U_0-U_{in,i}\right)\textrm{d}A.
\end{multline}
The second term on the right-hand side of (\ref{eq:zong3}) is not negligible for a turbine that experiences wakes of upwind turbines. However, this discrepancy does not seem to cause a noteworthy error in model predictions in the far wake. Figure \ref{Fig:stw_zong} shows variation of the centreline $(1-U_2/U_h)$ with downwind distance for a turbine that is subject to the full wake of another turbine. For a fair comparison, the same value of $k$ obtained from figure \ref{Fig:wake_recovery} is used for all results shown in figure \ref{Fig:stw_zong}. The figure shows that \citet{zong2020}'s model provides good far-wake predictions, which lie between results of the original and modified versions of conservation of momentum deficit. However, a drawback of this model could be the fact that it is not represented in an explicit form. To use this model, one needs to compute surface integrals of wake velocity deficit multiple times at each downstream location in order to obtain converged results through an iterative process. This may increase the computational cost of this superposition model compared with simpler superposition models as well as the explicit wind farm solution developed in the current study.

\color{black}
\section{Summary and future research}\label{sec:summary}
\color{black}
The main aim of this paper is to address the puzzling question of `\emph{how should one estimate cumulative wake effects in wind farms based on engineering wake models?}', given the abundance of wake superposition methods in the literature. \color{black}This aim is achieved by directly solving flow governing equations for a turbine that experiences upwind turbine wakes. The developed model can therefore predict the flow distribution in a wind farm without the need for any specific superposition method. \color{black} The LES data for wind farms with different number of rows are used to perform a budget analysis of the integral form of the mass and momentum equations for turbine wakes in wind farms. Results show that there are important non-negligible terms (e.g., pressure and Reynolds normal and shear stress terms) other than the momentum deficit and thrust force in this equation. However, the so-called conservation of momentum deficit (i.e., the equality of the momentum deficit term and thrust force) seems to be fairly valid in the far wake, at least for a moderately sized wind farm. A modified version of the conservation of momentum deficit is also introduced in an attempt to provide slightly better results at short downwind distances as well as in the far wake of turbines deep inside a wind farm. Performing LES with and without the presence of turbines allows us to properly examine some model assumptions. The data for both full-wake and partial-wake conditions show that wake velocity-deficit profiles are self-similar if they are defined with respect to the wind flow at the same position but in the absence of the turbine. While a Gaussian profile can successfully represent the self-similar profile of the wake at the centre, it falls short in capturing the flow speed-up at wake edges. This results in an underprediction of efficiency for certain turbines in primary rows that experience flow speed-up between adjacent columns.  

\color{black}  To quantify the wind farm flow distribution using this analytical model, the only necessary empirical input is the wake recovery rate. Provided that this is properly estimated for turbines within wind farms, our results show, overall, that the proposed model is able to provide acceptable predictions. Based on this model, the velocity distribution downwind of a wind farm consisting of $n$ wind turbines (WT$_1$,WT$_2$, ... WT$_n$) is given by
\begin{equation}\label{Eq:summary1}
U_{n}=U_{0}-\sum_{i=1}^{n}C_i \textrm{exp}\left( {-\frac{(y-y_i)^2+(z-z_i)^2}{2\sigma_i^2}}\right),
\end{equation}
where the wake half width $\sigma_i$ for WT$_i$ is estimated from (\ref{eq:sigma_i}). The value of $C_i$ (i.e., wake-centre velocity deficit associated with WT$_i$) is determined by
\begin{equation}\label{Eq:summary2}
\frac{C_i}{U_{h}}=\left(1-\sum_{j=1}^{i-1}\lambda_{ij}\frac{C_j}{U_{h}}\right)\left(1-\sqrt{1-\frac{c_{t,i}\left(\frac{<U_{i-1}>_{i,x_i}}{U_{h}}\right)^2}{8\left( \sigma_i/D \right)^2\left(1-\sum_{j=1}^{i-1}\lambda_{ij}\frac{C_j}{U_{h}}\right)^2}}\right),
\end{equation}
where $c_{t,i}$ is the thrust coefficient of WT$_i$. For $i=1$, the equation is reduced to the one for the single turbine wake by setting $\sum\lambda_{ij}C_j/U_h$ equal to zero. For $i> 1$, the value of $\lambda_{ij}$ in (\ref{Eq:summary2}) is given by 
\begin{equation}\label{Eq:summary3}
\lambda_{ij}=\frac{\alpha\sigma_j^2 \mathrm{e}^{-\frac{\left(y_i-y_j\right)^2}{2\left(\sigma_i^2+\sigma_j^2\right)}}
\mathrm{e}^{-\frac{\left(z_i-z_j\right)^2}{2\left(\sigma_i^2+\sigma_j^2\right)}}
}{\sigma_i^2+\sigma_j^2},
\end{equation}
where $\alpha$ is equal to 2 for the original equation of conservation of momentum deficit (\ref{Eq:momentum_deficit_wf}) and is equal to 1 for its modified version (\ref{Eq:lambda_modified}). \color{black} The value of $\lambda$ depends on wind farm turbine layout and inflow conditions. This coefficient, obtained analytically, functions similarly to empirical models (e.g., the `mosaic-tile' approach) that aim to quantify the net contribution of overlapping wakes. The validity of the common single turbine wake (STW) superposition models is also analysed in this study. It is shown that common superposition models can be derived by making approximations to the analytical wind farm flow solution presented in this work. 

\color{black}
There are still some important limitations that need to be better understood and addressed in future research if we want to successfully implement engineering wake models for a variety of inflow conditions and wind farm layout configurations:
\begin{itemize}
  \item[--] Neglected terms of the governing equation (\ref{Eq:Nav-Stk_general-x5}): The applicability of the so-called conservation of momentum deficit (\ref{Eq:momentum_deficit_wf}) (and its modified version (\ref{eq:modified_momentum deficit})) should be examined for wind farm arrays with various inflow conditions and layout geometries. Of special interest is to investigate the significance of different terms in (\ref{Eq:Nav-Stk_general-x5}) for a very large wind farm that asymptotes to a so-called fully developed case.
  
  \item[--] Turbine blockage effects: 
  One of the limitations of the proposed model is its inability to capture speed-up effects between adjacent turbine columns caused by turbine flow blockage. Future research can aim at using a more realistic relation (instead of Gaussian) to represent the self-similar velocity-deficit profiles. Furthermore, several recent studies  \citep[][]{bleeg2018blockage,segalini2020blockage} have demonstrated another important aspect of flow blockage in wind farms. These studies, among others, have shown that flow blockage caused by wind turbines within wind farms can decrease the efficiency of upstream turbines by reducing their effective incoming velocity. For more accurate prediction of power production in large wind farms, the effect of downwind turbines on their upwind counterparts needs to be included in wind farm flow engineering models. 
  \item[--] Wake recovery rate $k$: Model predictions are quite sensitive to the value of $k$. Therefore, accurate estimation and modelling of $k$ is of great importance. To achieve this goal, a better understanding of the turbulence distribution in wind farms and how this affects the turbine wake recovery rate is essential. Moreover, additional research is needed on the possible effects of any inflow and turbine operating conditions on wake recovery, other than the incoming turbulence intensity.

\end{itemize}
\color{black}
\vspace{5 mm}
{\bf Declaration of Interests}. The authors report no conflict of interest. 

\section*{Acknowledgements}
The authors thank S. Shamsoddin for his constructive feedback on the theoretical part of the work. The authors would also like to acknowledge Matthew J Churchfield from NREL for useful discussions regarding SOWFA.
A portion of the research was performed using computational resources sponsored by the U.S.~Department of Energy's Office of Energy Efficiency and Renewable Energy and located at the National Renewable Energy Laboratory.
This work was authored by the National Renewable Energy Laboratory, operated by Alliance for Sustainable Energy, LLC, for the U.S. Department of Energy (DOE) under Contract No. DE-AC36-08GO28308. Funding provided by the U.S. Department of Energy Office of Energy Efficiency and Renewable Energy Wind Energy Technologies Office. The views expressed in the article do not necessarily represent the views of the DOE or the U.S. Government. The U.S. Government retains and the publisher, by accepting the article for publication, acknowledges that the U.S. Government retains a nonexclusive, paid-up, irrevocable, worldwide license to publish or reproduce the published form of this work, or allow others to do so, for U.S. Government purposes.

\appendix

\bibliographystyle{jfm}
\bibliography{reference}

\end{document}